\documentclass[twocolumn,trackchanges]{aastex701}
\usepackage{amsmath}

\begin{document}

\title{Photometric Variability and Rotation of $\beta$~Pictoris~b from JWST NIRCam Coronagraphic Imaging
}

\correspondingauthor{Yifan Zhou}

\author[0000-0003-2969-6040]{Yifan Zhou}
\affiliation{Department of Astronomy, University of Virginia, Charlottesville, VA 22904, USA}
\email[show]{yzhou@virginia.edu}
\author[0000-0003-4614-7035]{Beth A. Biller}
\affiliation{Institute for Astronomy, University of Edinburgh, Royal Observatory, Blackford Hill, Edinburgh EH9 3HJ, UK}
\affiliation{Centre for Exoplanet Science, University of Edinburgh, Edinburgh, EH9 3HJ, UK}
\email{bbiller@ed.ac.uk}
\author[0000-0001-5365-4815]{Aarynn L. Carter}
\affiliation{Space Telescope Science Institute, 3700 San Martin Drive, Baltimore, MD 21218, USA}
\email{aacarter@stsci.edu}
\author[0000-0002-3191-8151]{Marshall D. Perrin}
\affiliation{Space Telescope Science Institute, 3700 San Martin Drive, Baltimore, MD 21218, USA}
\email{mperrin@stsci.edu}
\author[0000-0001-7739-9767]{Michael Poon}
\affiliation{Department of Astronomy and Astrophysics, University of Toronto, 50 St. George Street, Toronto, ON M5S 3H4, Canada}
\email{michaelkm.poon@mail.utoronto.ca}
\author[0000-0002-2011-4924]{Genaro Su\'{a}rez}
\affiliation{Department of Astrophysics, American Museum of Natural History, Central Park West at 79th Street, New York, NY 10024, USA}
\email{gsuarez@amnh.org}
\author[0000-0002-9962-132X]{Ben J. Sutlieff}
\affiliation{Institute for Astronomy, University of Edinburgh, Royal Observatory, Blackford Hill, Edinburgh EH9 3HJ, UK}
\affiliation{Centre for Exoplanet Science, University of Edinburgh, Edinburgh, EH9 3HJ, UK}
\email{ben.sutlieff@roe.ac.uk}
\author[0000-0003-0489-1528]{Johanna M. Vos}
\affiliation{School of Physics, Trinity College Dublin, Dublin 2, Ireland}
\email{johanna.vos@tcd.ie}
\author[0000-0003-0774-6502]{Jason J. Wang}
\affiliation{Center for Interdisciplinary Exploration and Research in Astrophysics (CIERA), Northwestern University, 1800 Sherman Avenue, Evanston, IL 60201, USA}
\affiliation{Department of Physics and Astronomy, Northwestern University, 2145 Sheridan Road, Evanston, IL 60208, USA}
\email{jason.wang@northwestern.edu}
\author[0000-0001-6396-8439]{William O. Balmer}
\affiliation{Department of Physics and Astronomy, Johns Hopkins University, 3400 N. Charles Street, Baltimore, MD 21218, USA}
\email{wbalmer1@jhu.edu}
\author[0000-0002-6076-5967]{Marta L. Bryan}
\affiliation{Department of Astronomy and Astrophysics, Penn State University, 525 Davey Laboratory, 251 Pollock Road, University Park, PA 16802, USA}
\affiliation{Department of Astronomy and Astrophysics, University of Toronto, 50 St. George Street, Toronto, ON M5S 3H4, Canada}
\email{martalbryan@psu.edu}
\author[0000-0001-9353-2724]{Anthony Boccaletti}
\affiliation{LESIA, Observatoire de Paris, Universit\'{e} PSL, CNRS, Sorbonne Universit\'{e}, 5 place Jules Janssen, 92195 Meudon, France}
\email{anthony.boccaletti@obspm.fr}

\author[0000-0001-8627-0404]{Julien H. Girard}
\affiliation{Space Telescope Science Institute, 3700 San Martin Drive, Baltimore, MD 21218, USA}
\email{jgirard@stsci.edu}

\author[0000-0003-4636-6676]{Eileen C. Gonzales}
\affiliation{Department of Physics and Astronomy, San Francisco State University, San Francisco, CA 94132, USA}
\email{egonzales@sfsu.edu}

\author[0000-0003-2769-0438]{Jens Kammerer}
\affiliation{European Southern Observatory, Karl-Schwarzschild-Str. 2, 85748 Garching, Germany}
\email{jens.kammerer@eso.org}
\author[0000-0002-0834-6140]{Jarron M. Leisenring}
\affiliation{Steward Observatory, University of Arizona, 933 N. Cherry Ave., Tucson, AZ 85721, USA}
\email{jarronl@arizona.edu}
\author[0000-0002-6217-6867]{Paulina. Palma-Bifani}
\affiliation{Laboratoire Lagrange, Universit\'{e} C\^{o}te d'Azur, CNRS, Observatoire de la C\^{o}te d'Azur, Nice, France}
\email{paulina.palma-bifani@oca.eu}
\author[0000-0002-4309-6343]{Kevin R. Wagner}
\affiliation{Steward Observatory, University of Arizona, 933 N. Cherry Ave., Tucson, AZ 85721, USA}
\email{kevinwagner@arizona.edu}
\author[0000-0003-3714-5855]{D\'aniel Apai}
\affiliation{Steward Observatory, University of Arizona, 933 N. Cherry Ave., Tucson, AZ 85721, USA}
\affiliation{Lunar and Planetary Laboratory, University of Arizona, 933 N. Cherry Ave., Tucson, AZ 85721, USA}
\affiliation{James C. Wyant College of Optical Sciences, University of Arizona, Tucson, AZ 85721, USA}
\affiliation{Earth, Atmospheric, and Planetary Sciences, Massachusetts Institute of Technology,  77 Massachusetts Avenue, Cambridge, MA 02139USA}
\email{apai@arizona.edu}
\author[0000-0001-5579-5339]{Mick\"ael Bonnefoy}
\affiliation{Univ. Grenoble Alpes, CNRS, IPAG, F-38000 Grenoble, France}
\email{mickael.bonnefoy@univ-grenoble-alpes.fr}
\author[0000-0003-2649-2288]{Brendan P. Bowler}
\affiliation{Department of Physics, University of California, Santa Barbara, Santa Barbara, CA 93106, USA}
\email{bpbowler@ucsb.edu}
\author[0000-0003-4557-414X]{Kyle Franson}
\affiliation{Department of Astronomy and Astrophysics, University of California, Santa Cruz, Santa Cruz, CA 95064, USA}
\email{kfranson@ucsc.edu}
\author[0000-0001-7047-0874]{Pengyu Liu}
\affiliation{Institute for Astronomy, University of Edinburgh, Royal Observatory, Blackford Hill, Edinburgh EH9 3HJ, UK}
\affiliation{Centre for Exoplanet Science, University of Edinburgh, Edinburgh, EH9 3HJ, UK}
\email{pengyu.liu@ed.ac.uk}
\author[0000-0001-8485-0325]{Marcio Mel\'endez}
\affiliation{Space Telescope Science Institute, 3700 San Martin Drive, Baltimore, MD 21218, USA}
\email{melendez@stsci.edu}
\author[0000-0003-3050-8203]{Stanimir A. Metchev}
\affiliation{Department of Physics and Astronomy, University of Western Ontario, London, ON N6A 3K7, Canada}
\email{smetchev@uwo.ca}
\author[0000-0003-0331-3654]{Simon Petrus}
\affiliation{NASA Goddard Space Flight Center, Greenbelt, MD 20771, USA}
\email{simon.petrus.pro@gmail.com}

\author[0000-0003-3818-408X]{Laurent Pueyo}
\affiliation{Space Telescope Science Institute, 3700 San Martin Drive, Baltimore, MD 21218, USA}
\email{pueyo@stsci.edu}
\author[0000-0002-4388-6417]{Isabel Rebollido}
\affiliation{Centro de Astrobiolog\'ia (CAB), CSIC-INTA, Camino Bajo del Castillo s/n, E-28692, Villanueva de la Ca\~{n}ada, Madrid, Spain}
\email{isabel.rebollidovazquez@esa.int}
\author[0000-0001-6098-3924]{Andrew J. Skemer}
\affiliation{Department of Astronomy and Astrophysics, University of California, Santa Cruz, Santa Cruz, CA 95064, USA}
\email{askemer@ucsc.edu}
\author[0000-0003-2278-6932]{Xianyu Tan}
\affiliation{Tsung-Dao Lee Institute, Shanghai Jiao Tong University, 520 Shengrong Road, Shanghai 201210, China}
\email{xianyut@sjtu.edu.cn}
\author[0000-0001-8818-1544]{Niall Whiteford}  
\affiliation{Department of Astrophysics, American Museum of Natural History, Central Park West at 79th Street, New York, NY 10024, USA}
\email{niallwhiteford@gmail.com}

\begin{abstract}
We report the detection of photometric variability in the directly imaged super-Jupiter $\beta$~Pictoris~b. Using JWST NIRCam dual-band coronagraphic imaging, we conducted a 16-hour continuous photometric monitoring campaign in the F210M and F410M filters. We developed and validated a time-series photometry framework that combines PSF subtraction, principal component analysis for systematic noise removal, and injection-and-recovery tests to confirm signal fidelity. Both light curves show consistent sinusoidal variability at ${\sim}5\sigma$ and $\gg 5\sigma$ significance in the F210M and F410M bands, respectively. A joint sinusoidal fit yields a rotation period of $P_{\rm rot} = 9.00 \pm 0.13$~hr and variability amplitudes of $0.85 \pm 0.07\%$ and $0.89 \pm 0.04\%$ in F210M and F410M, respectively. The near-identical amplitudes and periods in both bands confirm a common astrophysical origin in a heterogeneous atmosphere. Combining $P_{\rm rot}$ with the previously measured projected rotational velocity, we constrain the line-of-sight spin axis inclination of $\beta$~Pic~b. The result favors an equator-on viewing geometry, consistent with line-of-sight spin-orbit alignment: the planetary spin axis, orbital plane, debris disk, and stellar equator are all mutually aligned. This stands in sharp contrast to the large obliquities of wide-separation companions that are likely formed via gravitational fragmentation. Together with the system's young age, this observation provides independent dynamical evidence that $\beta$~Pic~b formed via core accretion. This result constitutes the first detection of rotational modulation in a close-in, high-contrast exoplanet that likely formed via core accretion, demonstrating that time-series coronagraphic imaging with JWST opens a powerful new window onto the rotation, atmospheric dynamics, and spin-orbit architecture of this population.
\end{abstract}


\section{Introduction}

Time-resolved observations of brown dwarfs and planetary-mass objects have revealed rotational modulations as the dominant driving mechanisms of photometric variability \citep[e.g.,][]{Scholz2005A&A...429.1007S,Scholz2015,Metchev2015,Vos2022ApJ...924...68V,Apai2021ApJ...906...64A,Fuda2024ApJ...965..182F}. These modulations arise from large, coherent atmospheric structures --- magnetic starspots in young and early-type objects \citep{Mohanty2003ApJ...583..451M,Reiners2008ApJ...684.1390R,Scholz2015}, and heterogeneous dust clouds in L and T dwarfs \citep{Lew2020AJ....159..125L,Tan2025}. The variability amplitude exceeds 20\% in $J$-band monitoring of the two most variable brown dwarfs \citep{Radigan2012ApJ...750..105R,Bowler2020ApJ...893L..30B}; for most monitored objects, the amplitudes range from sub-percent to a few percent levels \citep{Metchev2015,Vos2022ApJ...924...68V}. The prevalence of these signals has been well-established in isolated brown dwarfs and wide-separation companions; however, attempts to detect rotational modulations in close-separation directly imaged planets face significant challenges. Ground-based adaptive optics observations lack the photometric stability required to detect sub-3\% variability \citep{Apai2016ApJ...820...40A, Biller2021MNRAS.503..743B, Wang2022AJ....164..143W,Sutlieff2023MNRAS.520.4235S,Sutlieff2024MNRAS.531.2168S}. The \emph{Hubble Space Telescope} achieved high photometric stability but lacked sufficient high-contrast imaging capability at infrared wavelengths where these signals are strongest \citep{Zhou2016ApJ...818..176Z, Zhou2019AJ....157..128Z}.

Beyond atmospheric characterization, a measured rotation period ($P_\mathrm{rot}$) is the key ingredient for constraining planetary obliquity, which is the angle between the spin axis and the orbital plane. $P_\mathrm{rot}$, combined with the projected equatorial velocity $v\sin i$ from high-resolution spectroscopy and a radius estimate, constrains the spin-axis inclination \citep{Masuda2020AJ....159...81M}. Combined with the planet's orbit solution, the spin-axis inclination yields the planetary obliquity. Planetary obliquity is a direct diagnostic of formation and dynamical history \citep{Bryan2020AJ....159..181B, Bryan2021AJ....162..217B}. In our solar system, planetary obliquities span nearly the full range from Jupiter's near-zero tilt to Uranus' sideways rotation, each reflecting distinct dynamical histories \citep{Correia2001Natur.411..767C,Ward2004AJ....128.2501W}. The same mechanisms could operate in exoplanetary systems. Constraining an exoplanetary obliquity requires four elements: $P_{\rm rot}$ from photometric monitoring, $v\sin i$ from high-resolution spectroscopy, the orbital inclination from astrometry, and the planetary radius. Assembling all three for a single system is observationally demanding. Planetary obliquities have only been measured for four wide-orbit companions \citep{Bryan2020AJ....159..181B, Bryan2021AJ....162..217B, Palma-Bifani2023, Poon2024AJ....168..270P}. All show large misalignments, consistent with top-down formation in a gravito-turbulent environment \citep{Offner2016,Jennings2021MNRAS.507.5187J,Poon2025}.

$\beta$~Pictoris~b \citep[$\beta$ Pic b,][]{Lagrange2010Sci...329...57L} presents an exceptional opportunity for measuring planetary rotation and obliquity. The super-Jovian planet orbits within a well-studied debris disk system at 9--10~AU separation from its host star.   The system's angular momentum architecture is particularly well constrained. The stellar spin axis, the planetary orbital plane, and the debris disk orientation are all closely aligned, pointing to a primordially ordered configuration \citep{Kraus2020}. However, the warped secondary disk and non-zero planetary eccentricity suggest subsequent dynamical interactions \citep{Dawson2011ApJ...743L..17D}. Atmospheric retrievals indicate a heavy-metal mass of  $100\mbox{–}300\,M_\oplus$ within the planet, far exceeding the critical core mass for runaway accretion \citep{GRAVITYCollaboration2020, Wang2025}. Recent theoretical work proposes that such enrichment arises from major mergers between protoplanet cores, and that a Neptune-mass impact within the past 9–18 Myr could set $\beta$~Pic b ringing with seismic oscillations at percent-level amplitudes detectable with JWST \citep{Zanazzi2025}. High-resolution spectroscopy has measured the planet's projected rotational velocity \citep{Snellen2014Natur.509...63S,Parker2024MNRAS.531.2356P,Landman2024A&A...682A..48L}. Recent GRAVITY+ observations have additionally detected $^13$CO in the atmosphere of $\beta$ Pic b and reported tentative evidence for spectroscopic variability at a period of 4.4~hr \citep{vonStauffenberg2026}, further motivating time-resolved characterization of this planet. The precise rotation period is the last missing ingredient needed to complete the angular momentum architecture of this system. A spin-axis inclination consistent with $\sin i_p~\approx~1$ would confirm spin-orbit alignment and support formation via core accretion in a disk. A lower inclination would indicate dynamical disruption or primordial misalignment.

\newcommand{\bpicb}{\ensuremath{\beta}~Pic~b}

\bpicb{} has an early to mid-L spectral type (L3, $T_{\rm eff} = 1700$~K), and its spectral energy distribution suggests a dusty atmosphere  \citep{Bonnefoy2013A&A...555A.107B}. 
Young and low surface gravity objects are among the most variable substellar objects known \citep{Vos2018MNRAS.474.1041V, Vos2022ApJ...924...68V}. Detecting such modulations in $\beta$~Pic~b requires both high-contrast imaging capability and photometric precision beyond what ground-based AO instruments or HST can provide. JWST offers both. NIRCam coronagraphic observations exceed ground-based adaptive optics by an order of magnitude in photometric precision \citep{Carter2023, Rigby2023}. 

We present the first detection of rotational modulation in a high-contrast, close-in exoplanet that likely formed via core accretion. To achieve this, we conducted a 16-hour continuous monitoring campaign of $\beta$~Pic~b using JWST NIRCam dual-band coronagraphic imaging in the F210M and F410M filters (Program 4758, PI: Zhou, Figure~\ref{fig:observation}). These wavelengths probe the $K$ and $L$ band continuum, avoiding molecular absorption bands where modulation amplitudes are suppressed \citep{Apai2013,Zhou2018AJ....155..132Z}.

This paper is organized as follows. Section~2 describes the observations and data calibration. Section~3 details the time-series photometry procedures. Section~4 presents the validation of the variability detection. Section~5 discusses the rotation period, the origin of the variability, the planetary obliquity, and the implications for JWST time-series high-contrast imaging. Section~6 summarizes our conclusions.

\section{Observations and data calibration}

\subsection{Observations} \label{sec:observations}

\begin{figure*}

    \centering
    \includegraphics[width=1\linewidth]{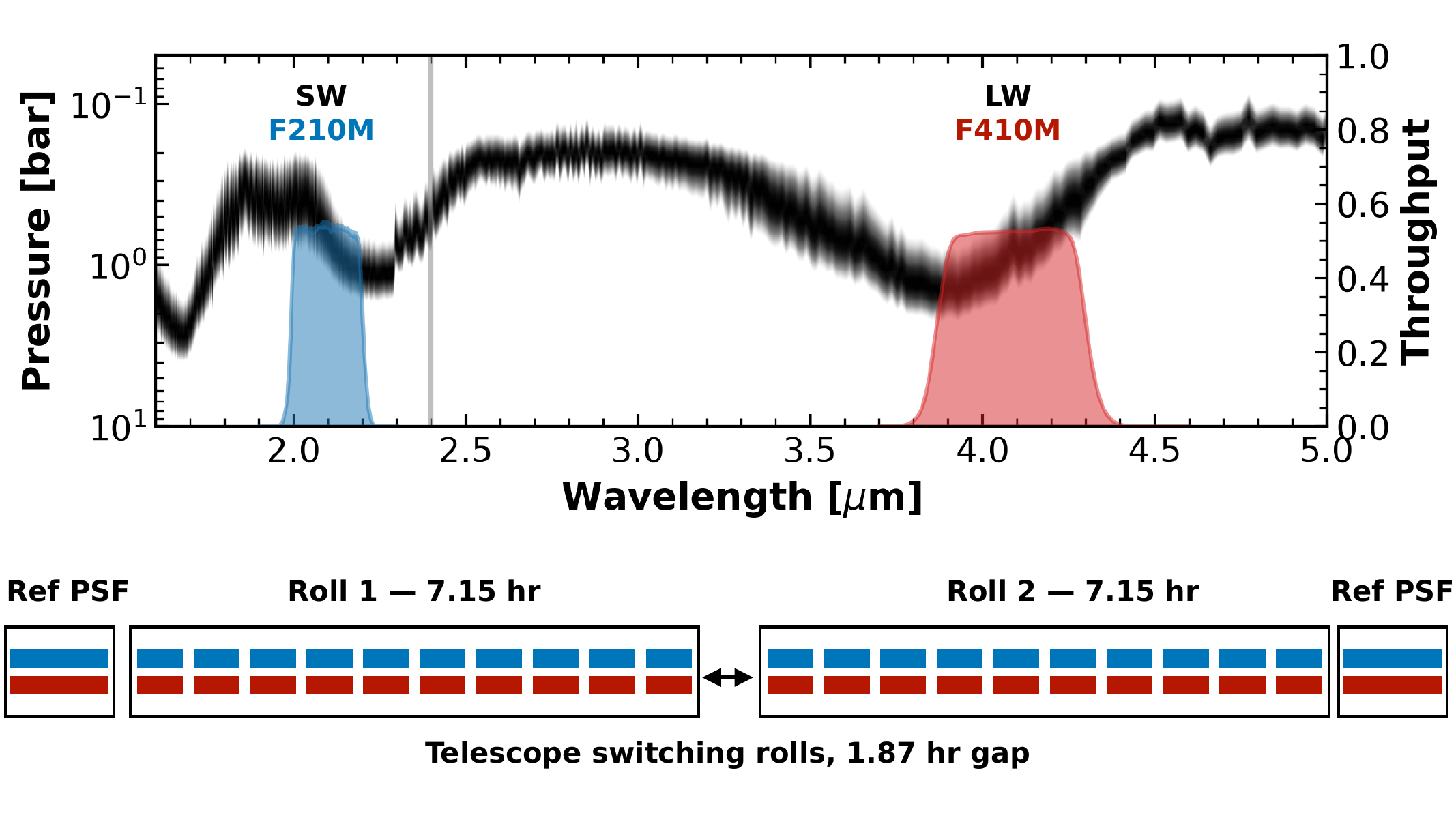}
    \caption{Filter selection and observation sequence for the $\beta$~Pic~b monitoring campaign. Top: transmission functions of the F210M (blue) and F410M (orange) filters overlaid on the probed atmospheric pressure of $\beta$~Pic~b (black). The probed pressure curve is derived assuming a $T_\mathrm{eff}=1700\,$K cloudy Exo-Rem model \citep{Charnay2018}. Both filters probe the $K$ and $L$ band continuum at similar photospheric pressure levels. Bottom: observation sequence comprising two telescope rolls of 7.15~hr each, separated by a 1.87~hr gap, with simultaneous F210M (blue) and F410M (red) exposures.}
    \label{fig:observation}
\end{figure*}

We observed $\beta$ Pictoris b with JWST NIRCam coronagraphic imaging on March 21--22, 2025 (Program 4758, PI: Zhou). Figure~\ref{fig:observation} illustrates the design of the observing campaign. The observations comprised 20 non-dithered consecutive exposures of $\beta$ Pic split into two rolls of ten exposures each. Two identical sets of reference star observations of $\alpha$ Pic bracketed the science sequence. All exposures employed dual-band coronagraphic imaging with the MASK335R occulter, F210M ($\lambda_\mathrm{cen}=2.10\,\mu$m, FWHM$=0.21\,\mu$m) in the short-wavelength channel, and F410M ($\lambda_\mathrm{cen}=4.08\,\mu$m, FWHM$=0.44\,\mu$m) in the long-wavelength channel. These filters probe the $K$ and $L$ band continuum and avoid molecular absorption.

The coronagraphic science exposures achieved a time baseline of 16.2 hours, spanning from the first science exposure in Roll 1 at 11:23:20 UT on March 21 to the last exposure in Roll 2 ending at 03:34:07 UT on March 22. Additional time was spent on guide star acquisition, coronagraphic target acquisition, and other setup at the start of Roll 1. The position angle rotated by 9.9 degrees between the two rolls. A 112-minute (1.87 hour) gap in coverage separated the science observations in the two rolls\footnote{This gap was longer than intended due to a side effect of JWST's event driven operations. An earlier observation had failed, opening up a gap in the timeline, and the onboard software was able to start Roll 1 about 2.7 hours earlier than scheduled. Operational timing constraints on Roll 2 limited it to starting no more than 1.6 hours earlier than scheduled. This difference opened up an unintended gap of 1.1 hours after roll 1 finished but before roll 2 could start, during which JWST was idle.  That 1.1 hour gap sums with the $\sim 0.75$ hour duration of the slew, guide star acquisition, and coronagraphic target acquisition at the start of roll 2 to produce the total gap between the science exposures in the two rolls. }. Each science exposure consisted of 110 integrations with the BRIGHT2 readout pattern and 10 groups per integration. This configuration yielded an exposure time of 2352 seconds and a cadence of 22.5 seconds between successive integrations. This was repeated 10 times in each roll for a total of 7.15 hr per roll. 

The reference star observations used the BRIGHT2 readout pattern with 5 groups and 50 integrations per exposure, repeated for 9 dithers. The nine-point small-grid dither pattern improves the spatial sampling of the reference PSF frames, a proven technique to mitigate target acquisition residuals and thereby improve high-contrast coronagraphic imaging \citep{Carter2023,Kammerer2022,Kammerer2024}.  The reference star, $\alpha$ Pic, is about 1 magnitude brighter than $\beta$ Pic, so the lower number of groups yields a comparable illumination level per integration, and thus comparable fraction of detector pixel well depth filled per integration. 

\subsection{Imaging data calibration}

We reduced the observations using the \texttt{spaceKLIP} package \citep{Carter2025}. The reduction started with \texttt{uncal} frames downloaded from the STScI MAST archive. We performed \texttt{stage1} and \texttt{stage2} reductions to acquire flux-calibrated \texttt{cal} and \texttt{calints} files. For the up-the-ramp fitting step, we tested both the default \texttt{OLS} ramp-fitting algorithm and the newly implemented \texttt{LIKELY} algorithm \cite{Brandt2024a,Brandt2024b}, and verified that both methods yield fully consistent calibrations.

Accurate bad pixel identification and correction is critical for high-precision time-series photometry. By default, \texttt{spaceKLIP} identifies and corrects bad pixels at the exposure level. Incorrectly identified or over-corrected bad pixels introduce artifacts compromising all integrations in the exposure, which ultimately result in inaccurate photometric measurements. We identify and correct bad pixels in four steps. First, we flag pixels marked with ``DO NOT USE'' and ``NON SCIENCE'' flags in the data quality array. Second, we flag pixels from a custom bad pixel list that includes manually identified outliers in the median frame. Pixels identified in these first two steps are replaced with the median of neighboring good pixels in a $5 \times 5$ box. Third, we apply the \texttt{timeints} method implemented in \texttt{spaceKLIP}, which identifies outliers in the temporal dimension. We adopt a conservative $20\sigma$ threshold. The identified pixels are replaced using the \texttt{timemed} method, which substitutes the median value of good pixels within the same exposure. Fourth, we apply the \texttt{sigclip} method implemented in \texttt{spaceKLIP}, which uses an iterative sigma-clipping algorithm to identify outliers in the spatial dimension. The detection threshold is also $20\sigma$. These pixels are replaced by the median of neighboring good pixels in a $5\,\mbox{pixels} \times 5\,\mbox{pixels}$ box. This procedure produces smooth light curves without spurious or outlier values across the image.

\begin{figure}[!t]
    \centering
    \includegraphics[width=1.0\linewidth]{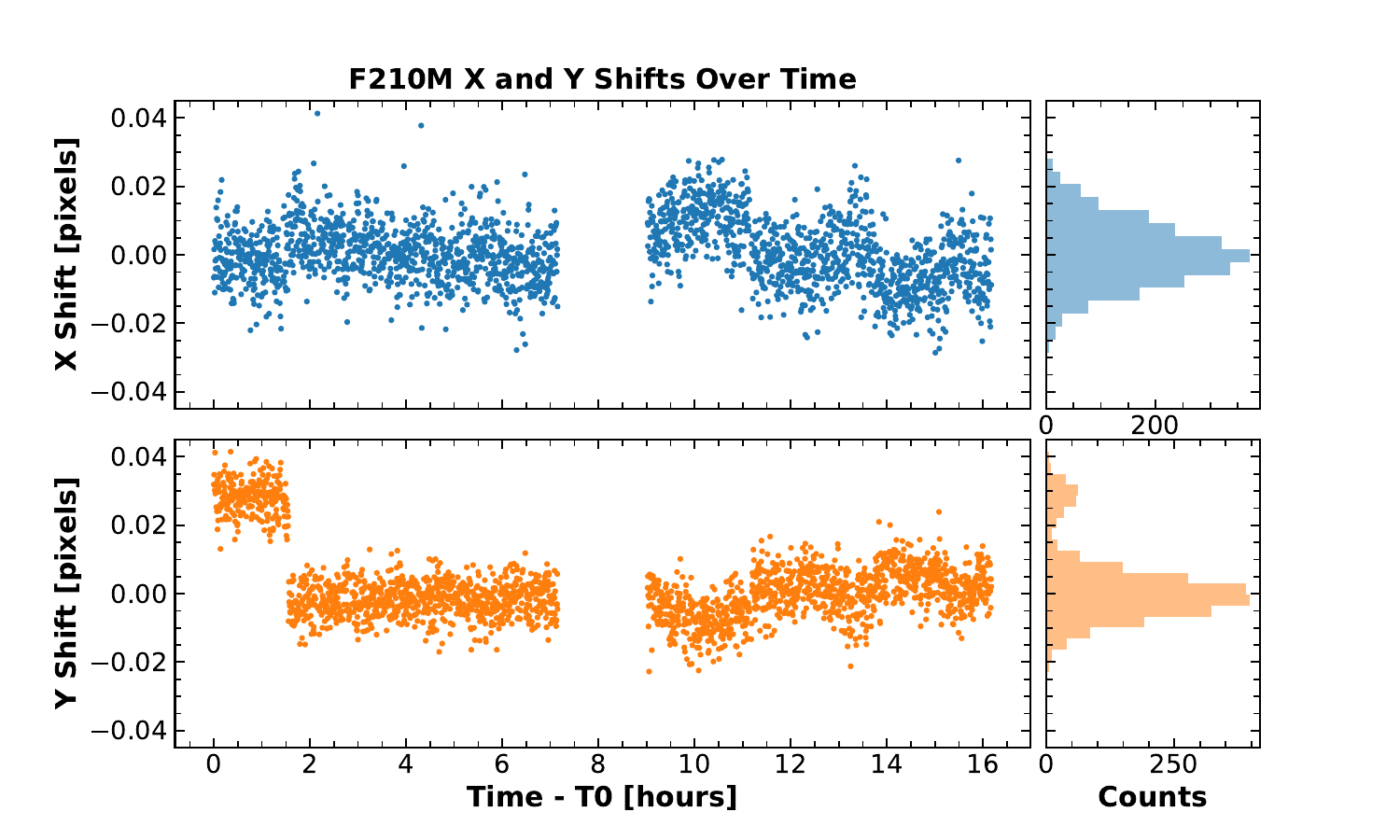}
    \caption{Centroid shifts relative to the median value of either rolls in $x$ and $y$ directions. The telescope maintained extremely precise pointing ($\sigma\sim0.01$\,pixels in F210M, 0.3~mas) during the observations. A pointing jump occurred at 1.55 hr after the start of the first science integration due to a mirror tilt event then.}
    \label{fig:alignment}
\end{figure}

We aligned the images by the stellar positions found by the \texttt{calculate\_centers} function implemented in \texttt{spaceKLIP}. It determines the stellar position by fitting a \texttt{stpsf} \citep{2012SPIE.8442E..3DP, 2014SPIE.9143E..3XP} model to each frame. The \texttt{stpsf} model was generated using wavefront information retrieved from MAST via the \texttt{load\_wss\_opd\_by\_date} function implemented by the \texttt{stpsf} package. We provided an A5 spectral template to \texttt{STPSF} to match the spectral type of $\beta$ Pic, minimizing chromatic differences between the model PSF and the observations. The measured centroid offsets ($\Delta x$ and $\Delta y$) were normally distributed with a standard deviation of (0.009, 0.011) pixels in the F210M band and (0.006, 0.004) pixels in the F410M band (Figure~\ref{fig:alignment}). 

A noticeable discontinuity occurred approximately 1.5 hours into Roll 1 at 2025-03-21 12:56, which can be seen as changes in both the apparent position and brightness of the star as seen in the NIRCam data. We initially hypothesized this might be due to an unintended guide star motion or guiding issue, but eventually we determined it was due to a small ``tilt event'' mirror motion then. Tilt events are a known but unpredictable occasional phenomenon in JWST's optics, due to the release of stored structural stresses from the observatory's initial cooling \citep{2024SPIE13092E..10T}. The longer duration of time series observations gives more opportunity for tilt events to occur, and such events have been seen in previous TSO observations  \citep[e.g.][]{2023PASP..135a8001S, 2023PASP..135g5001A, 2023Natur.614..664A, Chen2025}. Similar to those prior works, we treat the resulting discontinuity as a systematic to be removed via data processing steps described in Section~\ref{sec:reduction}

We provide further diagnostic information on the tilt event in Appendix \ref{app:tilt_event}. Notably, at the exact time of the discontinuity seen in the NIRCam coronagraphic data, a transient bright flash was detected in the JWST guider data, which is a known signature of some micrometeorite impacts into the observatory. This dataset provides the strongest evidence yet supporting the hypothesis that at least some tilt events may be triggered by micrometeoroid impacts onto non-optical surfaces.

We aligned all science and reference integrations to the first science integration using the \texttt{spaceKLIP} \texttt{calculate\_alignment} and \texttt{shift\_image} methods. Before shifting, we low-pass filtered each image with a Gaussian kernel (FWHM = 0.4 times the PSF FWHM). This eliminated bandwidth-limit artifacts due to spatial undersampling in the aligned images. The shifting was performed in the Fourier domain using the \texttt{fourier\_shift} algorithm implemented in \texttt{spaceKLIP}.

\section{Time-resolved coronagraphic imaging with NIRCam -- data reduction strategy}
\label{sec:reduction}

We analyze the F210M and F410M data independently. Each analysis follows three steps. First, we measure time-resolved aperture photometry centered on the host star. This photometry characterizes and calibrates stellar variability. Second, we perform PSF subtraction and forward modeling. This step determines the best planet model and the roll-averaged planet photometry. Third, we measure aperture photometry centered on the planet and on comparative speckle regions. This photometry yields the planet light curve. We detail each step in the following sections.

\subsection{Host star light curve characterization}

\begin{figure*}[!t]
    \centering
    \includegraphics[width=1\linewidth]{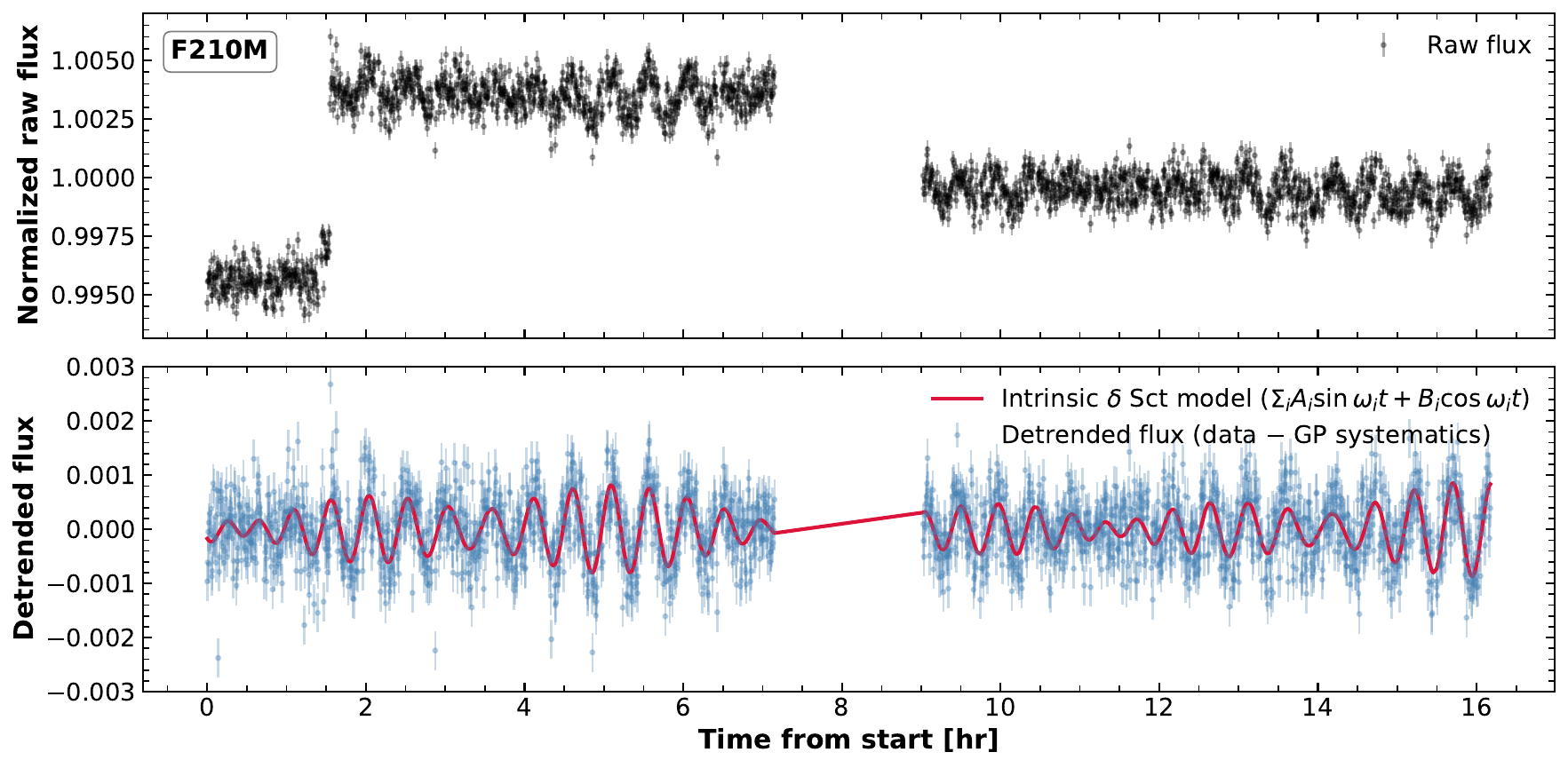}
    \caption{F210M light curve of the host star. \textit{Upper panel:} The raw aperture photometry shows strong oscillation signals consistent with $\beta$ Pic's known oscillation periods. Instrument systematics induce apparent flux jumps. \textit{Lower panel:} A Gaussian process successfully removes the systematic trend. The corrected light curve is well fit by four sine waves representing the strongest oscillation modes (red curve).}
    \label{fig:stellarLC}
\end{figure*}

$\beta$ Pic is a $\delta$-Scuti variable and its oscillation mode periods have been characterized in detail by ground- and space-based monitoring campaigns \citep{Mekarnia2017,Kenworthy2021,Zieba2019}. We model this variability to ensure stellar oscillations do not compromise the planet light curve measurements. 

The raw light curves are measured using a circular aperture with radius $r = 30$ pixels centered on the star. Starlight diffracted by the coronagraph dominates the flux within this aperture. Planet and debris disk emission are negligible \citep{Kammerer2024}. The stellar oscillation is apparent in the raw light curve. The amplitude is low (less than 0.1\%), and the frequency matches the known main oscillation mode. Instrument systematics also appear in the light curve. The flux shows a jump at 1.55 hr, associated with the tilt event, and another jump between the two telescope rolls which is plausibly due to slight differences in target acquisition performance on the star between the two rolls.

We model these systematics with a Gaussian process using the package \texttt{george} and a three-dimensional Matern~3/2 kernel. The first two dimensions are the $x$ and $y$ spatial coordinates. These dimensions decorrelate the light curve from pointing drift. The third dimension is time. This dimension characterizes remaining correlated noise in the time domain.

We combine the Gaussian process with a physical model for stellar oscillation-induced variability. The physical model captures the four strongest oscillation modes identified by previous observations ($P = 0.5049, 0.5284, 0.4835$, and 0.4424\,hr, \citealt{Zieba2019}). Adding additional oscillation modes does not improve the fit. We model each mode $A\sin(\omega t) + B\cos(\omega t)$ ($A$ and $B$ are free parameters). The joint model includes three hyperparameters characterizing the Gaussian process and a total of eight sine wave parameters. We fit this model to the observed light curve using the least squares method implemented by the \texttt{lmfit} package \citep{Newville2016}. The model describes the data well (Figure~\ref{fig:stellarLC}). The amplitudes are less than 0.02\% (200 ppm) in both bands. We save the intrinsic stellar light curve model for subsequent analysis.

This strategy is effective because the stellar oscillation modes have high signal-to-noise ratios. Low signal-to-noise signals would likely be eliminated through overfitting of the systematic model.

\begin{figure*}[!t]
    \centering
    \includegraphics[width=1.0\linewidth]{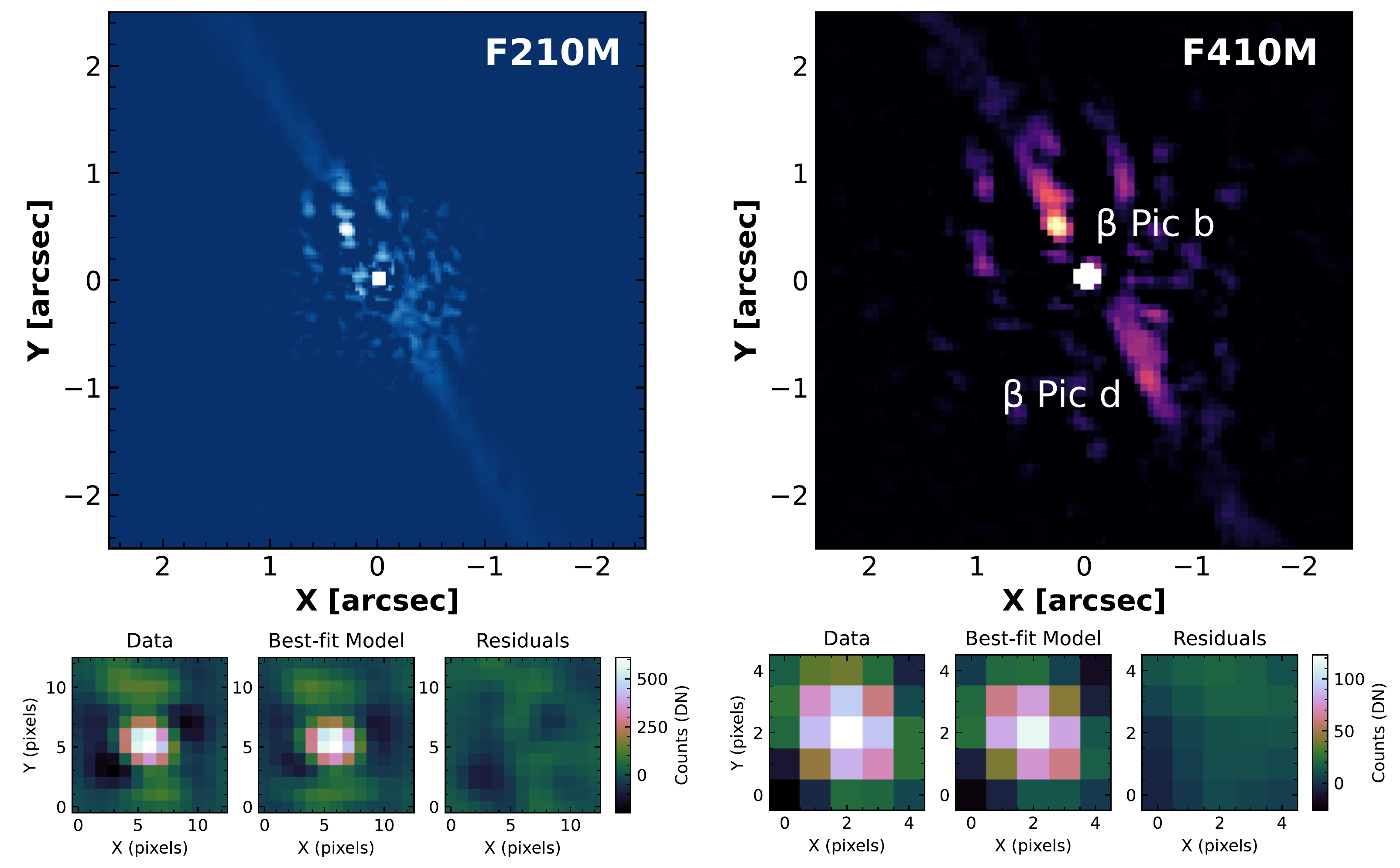}
    \caption{\texttt{spaceKLIP} PSF subtraction yields high S/N detection of $\beta$ Pic b in the F210M and F410M bands (upper left and upper right). The newly discovered planet $\beta$ Pic d \citep{Gibbs2026, Sutlieff2026} is visible in the F410M band image. KLIP Forward models using \texttt{STPSF} models successfully describe the detected planet signal and provide precise time-averaged photometry estimates (lower panels).}
    \label{fig:subtraction_fm}
\end{figure*}

\subsection{PSF subtraction and forward modeling photometry}

We apply PSF subtraction and forward modeling on the entire data set to determine the best-fitting time-averaged PSF model for $\beta$ Pic b. We process the F210M and F410M data using the same procedure, with filter-specific parameters noted below. The analysis uses coadded images. Each coadded image combines all 
integrations within one exposure. This approach works because the JWST PSF is stable across the entire observing sequence. Previous JWST NIRCam observations of $\beta$ Pic b verified this method \citep{Kammerer2024}.

Even with the tilt event that occurred part way through Roll 1, the PSF is still sufficiently stable that this is a practical approach. In particular the location of the affected mirror segment and the shape of the NIRCam coronagraphic Lyot stop result in, for this particular tilt event, mostly an apparent slight position shift rather than any substantial change in the coronagraphic PSF speckle pattern (Appendix \ref{app:tilt_event}). Empirically a single time-averaged PSF model worked sufficiently well for the bulk starlight subtraction, and we opt to handle the remaining small systematic offsets from the tilt event subsequently at the light curve stage as described below.  A high-pass filter with a 5-pixel wide window is applied to each image before PSF subtraction. This filter removes low spatial frequency variations introduced by scattered light from the disk \citep{Kammerer2024}.

The PSF subtraction is performed using \texttt{spaceKLIP} with combined Reference-star Differential Imaging \citep[RDI,][]{2009ApJ...694L.148L} from the dithered $\alpha$~Pic images and Angular Differential Imaging \citep[ADI,][]{2006ApJ...641..556M}, using 50 Karhunen-Loève Image Projection \citep[KLIP,][]{2012ApJ...755L..28S,Pueyo2016} basis vectors. The KLIP reduction divides the field into three annular zones: an inner zone, a dedicated planet zone spanning from 10 pixels inside to 15 pixels outside the planet separation, and an outer zone. KLIP is performed on these three zones independently. Figure~\ref{fig:subtraction_fm} (upper panels) shows the PSF-subtracted images.

KLIP forward modeling constrains the planet flux and position \citep{Pueyo2016}. The PSF models are generated using \texttt{stpsf}, incorporating the observation date, the closest available wavefront estimates, the telescope roll angle, and the estimated planet-star separation based on the astrometry solution from \citet{Lacour2021}. The \texttt{pyKLIP} FM module then uses these PSF models to construct and evaluate the KLIP forward model. A single \texttt{stpsf} model is applied to all exposures; as noted above, this is a simplification but is sufficient since the tilt event did not substantially change the speckle pattern. The same 5-pixel high-pass filter is applied to the PSF model for consistency with the data. MCMC optimization determines the best-fitting planet position and flux using 100 walkers, 500 burn-in steps, and 1000 sampling steps. The fitting is performed within a stamp of 13 pixels for F210M and 5 pixels for F410M, centered on the planet. The forward model successfully reproduces the planet signal (Figure~\ref{fig:subtraction_fm}, lower panels). PSF subtraction delivers high-quality detections of the planet in both bands.  The signal-to-noise ratio, estimated as the flux divided by the 1$\sigma$ flux uncertainty, is 61.2 in F210M and 39.5 in F410M.

The MCMC best-fit planet position and flux are then used to compute the pixel-by-pixel planet flux contribution in each science image. For each exposure, the forward-modeled planet PSF, scaled to the best-fit flux and oriented to the corresponding roll angle, is divided by the observed image to yield the fractional planet contribution at each pixel. This contribution map defines the planet flux fraction $c_i$ used in the aperture photometry described in the following section.

\subsection{Time series photometry of \bpicb}


\begin{figure}[!ht]
    \centering
    \includegraphics[width=1.0\linewidth]{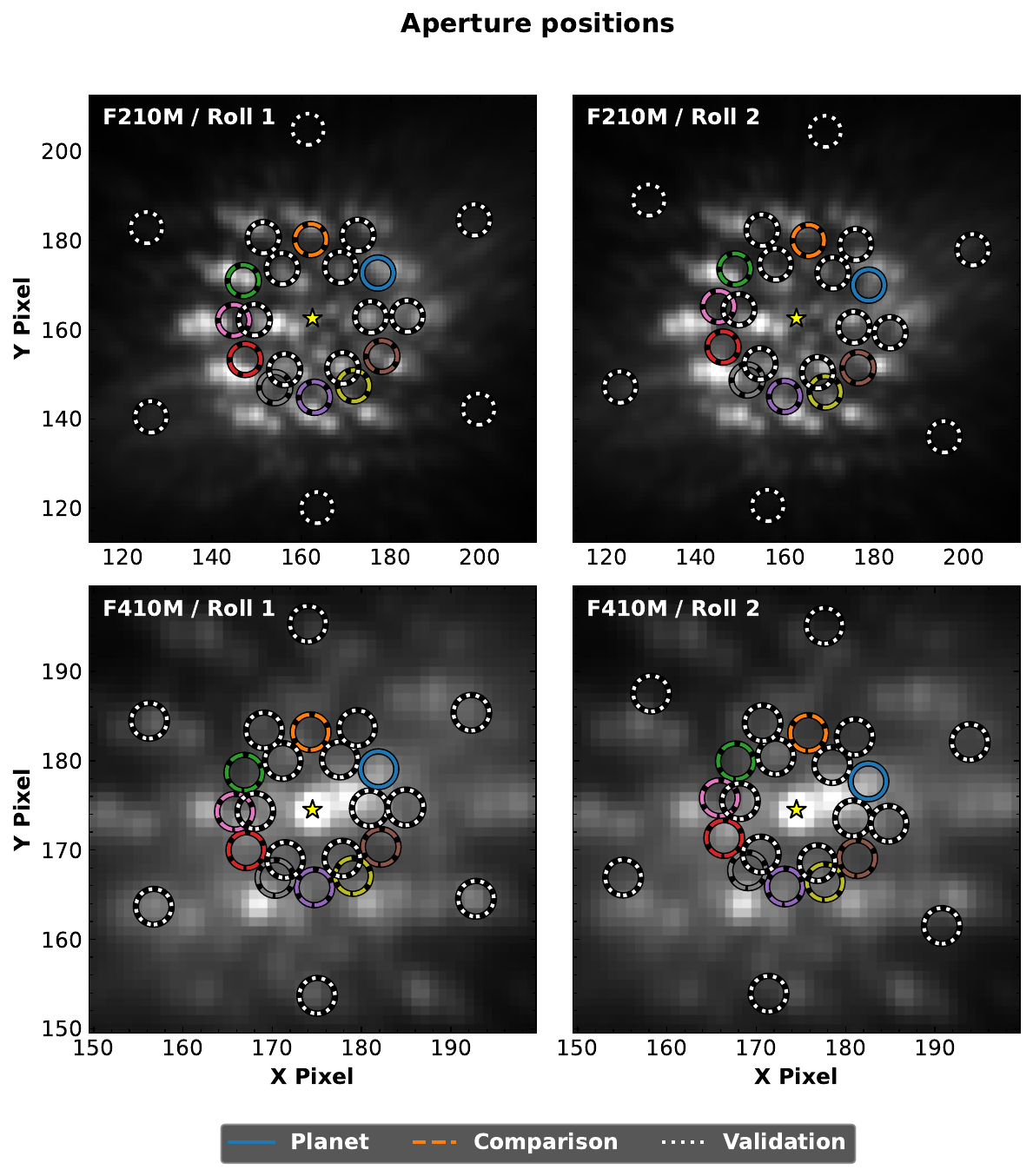}
    \caption{The aperture positions. Apertures for $\beta$ Pic b, comparison positions, and validation positions are shown in blue solid, color dashed, and white dotted circles, respectively. Upper panels are for the F210M band and lower panels are for the F410M band. Nominal aperture radii of 3.5 pixels for F210M and 2 pixels for F410M are represented by the size of the circles.}
    \label{fig:apertures}
\end{figure}

Time-resolved aperture photometry extracts the light curve from centroid-aligned \texttt{calints} images. These images are not primary subtracted. We account for the primary star flux as part of our light curve models. The analysis isolates the planet signal from systematic noise and stellar variability.

\begin{figure*}[!ht]
    \centering
    \includegraphics[width=1.0\linewidth]{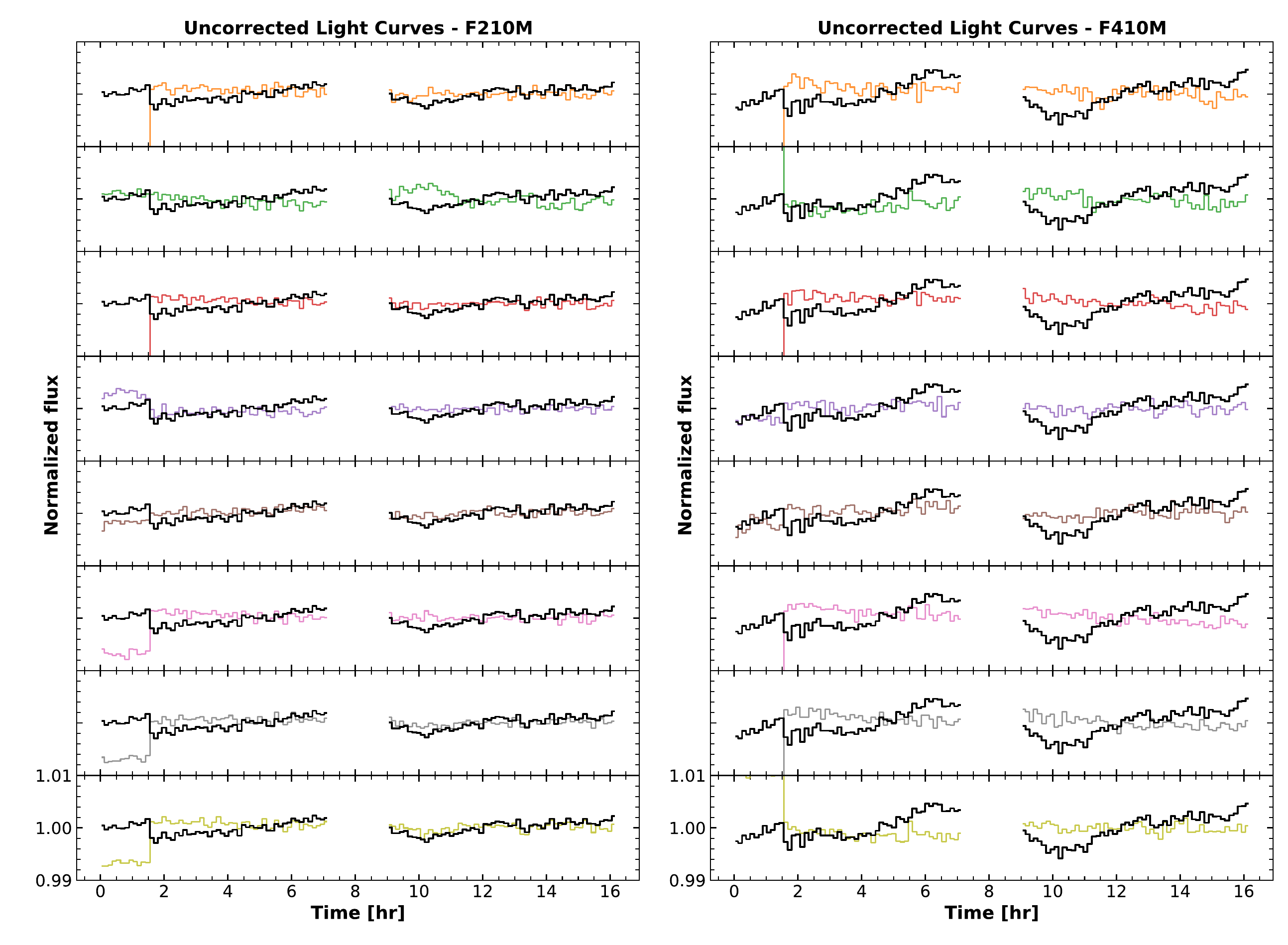}
    \caption{F210M and F410M light curves (the $F_i(t)$ term in Equation~\ref{eq:S1}). Light curves measured in the planet apertures are plotted in black lines and the comparison aperture light curves are plotted in the same colors as their apertures in Figure~\ref{fig:apertures}.}
    \label{fig:rawlightcurve}
\end{figure*}

Aperture photometry measures flux in circular apertures centered on the planet location. The planet position in each telescope roll is taken from the forward modeling results. Comparison apertures are placed at the same angular separation but different position angles. The locations of these apertures are shown in Figure~\ref{fig:apertures}. Flux in the comparison apertures is dominated by diffracted stellar and debris disk emission. For aperture radii, we use $r=3.5$ pixels ($0.11''$, $1.6\times$~FWHM) for the F210M filter and $r=2.0$ pixels ($0.13''$, $0.95\times$~FWHM) for the F410M filter. The choice of aperture size (between 3 to 4 pixels for F210M; between 1.5 to 2.5 pixels for F410M) does not impact the relative photometry results. Figure~\ref{fig:rawlightcurve} demonstrates the extracted raw light curves in the planet aperture and the comparison apertures.

\begin{figure*}
    \centering
    \includegraphics[width=1\linewidth]{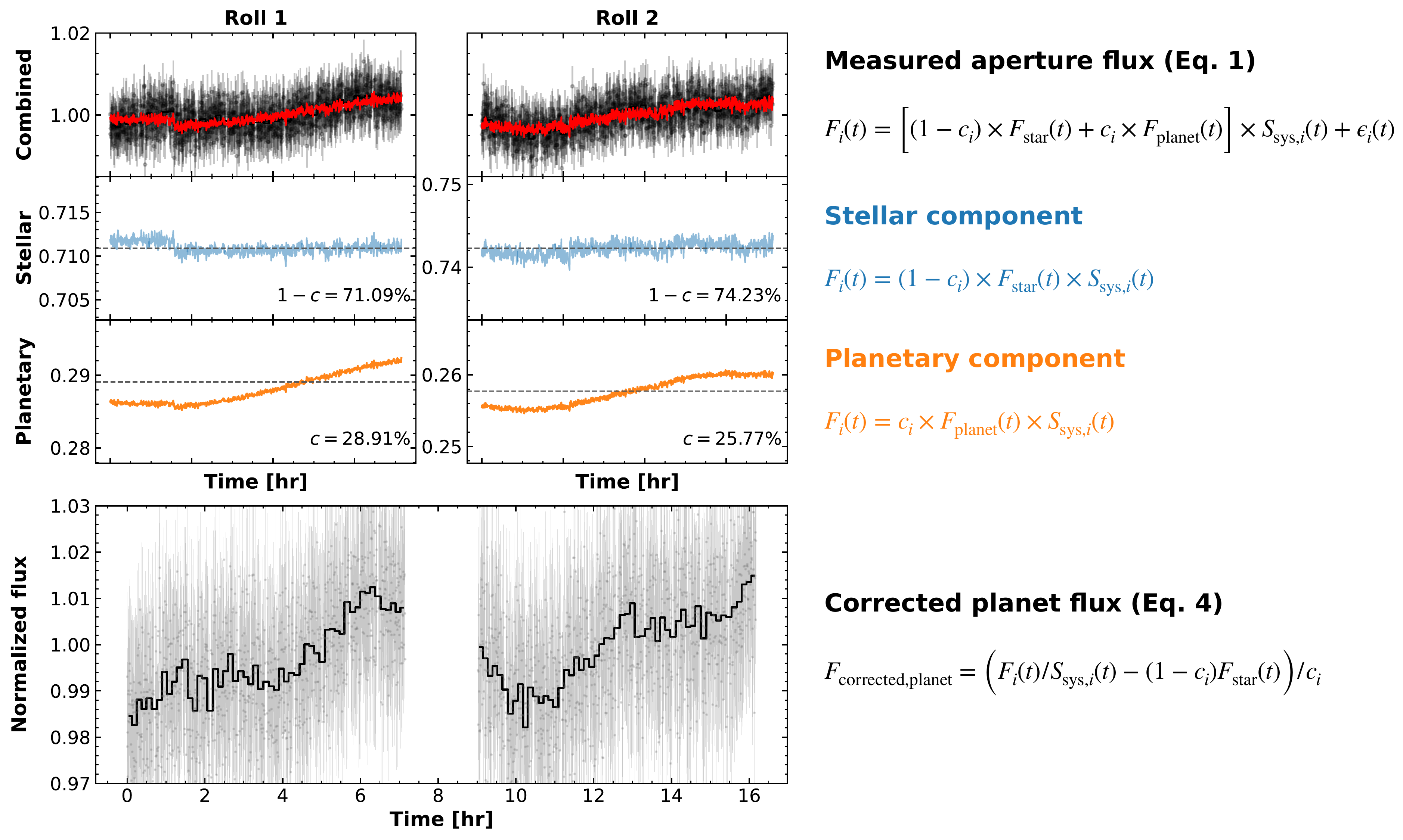}
    \caption{Demonstration of the light curve correction procedure using the F410M data. The top panels show the measured planet-aperture light curves in the two telescope rolls. Black dots are the raw flux measurements. Red lines are the best-fitting light curve model (Eq.~\ref{eq:S1}). The second and third rows decompose the aperture model into stellar (blue) and planetary (orange) components. The bottom row shows the isolated planetary light curve. The corresponding correction formula are provided on the right.}
    \label{fig:correction_demo}
\end{figure*}

The measured flux in each aperture contains contributions from the planet, diffracted starlight, and disk light. Time-dependent instrumental systematics further distort the light curve. A linear model describes the observed normalized light curve $F_i(t)$ for the $i$th aperture:
\begin{equation}
\begin{split}
F_i(t) &= \left[(1-c_i) \times F_{\mathrm{star}}(t)+ c_i \times F_{\mathrm{planet}}(t)\right]\\
&\times S_{\mathrm{sys}, i}(t) + \epsilon_i(t), \label{eq:S1}    
\end{split}
\end{equation}
where $c_i$ is the visit-average planet flux contribution fraction, $F_{\mathrm{star}}(t)$ is the stellar light curve model (fixed, Figure~\ref{fig:stellarLC}), $F_{\mathrm{planet}}(t)$ is the planet light curve model, $S_{\mathrm{sys}, i}(t)$ is the systematic trend in the $i$th aperture, and $\epsilon_i(t)$ is the random measurement error. We calculate the visit-average planet contribution factor by integrating the forward-modeled planet PSF within the aperture boundaries. The top row in Figure~\ref{fig:correction_demo} demonstrates this term.

The systematic term $S_{\mathrm{sys}, i}(t)$ describes any trends introduced by telescope pointing shifts or other instrumental effects. We empirically determine this term using light curves integrated from the comparison apertures. These apertures contain negligible planet flux ($c_i \approx 0$), so the observed flux reduces to $F_i(t) = F_{\mathrm{star}}(t) \times S_{\mathrm{sys},i}(t) + \epsilon_i(t)$. In principle, the systematic term should be estimated as $S_{\mathrm{sys},i}(t) = [F_i(t) - \epsilon_i(t)] / F_{\mathrm{star}}(t)$, but $\epsilon_i(t)$ is a realization of random noise and is not separable from the signal. We therefore use $S_{\mathrm{sys},i}(t) = F_i(t) / F_{\mathrm{star}}(t)$ as the empirical estimate, which folds the random error into the systematic term. The residual random noise sets a noise floor of 0.3\% in F210M and 0.15\% in F410M on the systematic model estimate.

Principal component analysis (PCA) identifies the dominant trends in systematic noise while avoiding overfitting. We construct a data matrix from the light curves of all comparison apertures. PCA decomposes this matrix to reveal the modes of correlated variability that represent systematic effects common across all apertures. In both the F210M and F410M bands, the first three principal components explain over 90\% of the total variance. For the first roll, the first principal component alone explains 99\% of the total variance. We use the first three principal components to model the systematic error in the light curve and express the systematic term as:
\begin{equation}
S_{\mathrm{sys}} = w_0 + w_x \times \mathrm{PC}_x + w_y \times \mathrm{PC}_y + \sum_{i=1}^{3} w_i \times \mathrm{PC}_i 
\label{eq:systematics}
\end{equation}
where $w_0$ is a normalization constant, $\mathrm{PC}_x$ and $\mathrm{PC}_y$ are the normalized stellar centroid shifts in the $x$ and $y$ directions, defined as the per-frame centroid displacement minus the median displacement (Figure~\ref{fig:alignment}) divided by its standard deviation, $w_x$ and $w_y$ are their respective weights, and $w_i$ are the weights for each principal component. In Figure~\ref{fig:correction_demo}, the second and third row panels isolate the stellar and planetary light curve components.

Rotational variability in brown dwarfs and directly imaged planets at similar 
wavelengths often manifests as sinusoidal modulation at the wavelengths used in this study \citep[e.g.,][]{Metchev2015, Vos2022ApJ...924...68V}. 
We adopt this as a nominal model for the planetary signal:
\begin{equation}
\label{eq:planet}
F_{\mathrm{planet}} = 1 + A_p \sin\left(\frac{2\pi t}{P_p}\right) + B_p 
\cos\left(\frac{2\pi t}{P_p}\right),
\end{equation}
which is described by the amplitudes $A_p$ and $B_p$, and period $P_p$. The 
impact of this model assumption on signal detection is evaluated in 
Section~\ref{sec:model_impact}. With six parameters characterizing the systematic model ---the normalization $w_0$, the centroid weights $w_x$ and $w_y$, and three principal component weights $w_1$ through $w_3$ --- each light curve is described by nine free parameters in total. We fit the model expressed by Equation~\ref{eq:S1} to the light curves extracted from the planetary aperture and the comparison apertures to obtain the best-fitting systematic model weightings and the planetary model parameters. Fitting for each light curve in individual telescope roll is performed separately using a Markov Chain Monte Carlo (100 walkers, 500 burn-in steps, and 1000 sampling steps).

\begin{figure*}[!th]
    \centering
    \includegraphics[width=1\linewidth]{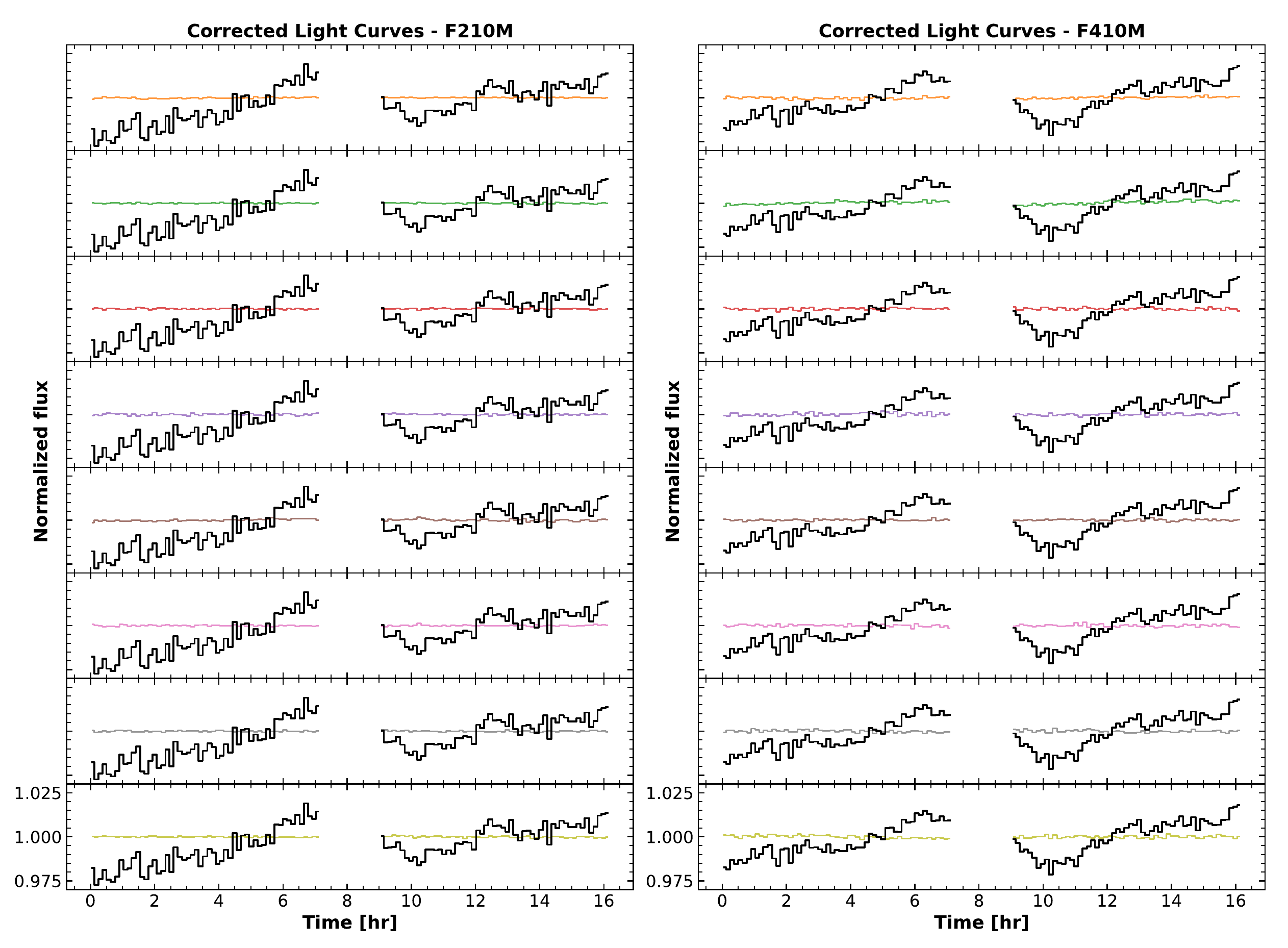}
    \caption{Light curves after the PCA model correction in F210M (left) and F410M (right).The corrected light curves of $\beta$ Pic b (black solid lines) are compared with those extracted from the comparison apertures.}
    \label{fig:correctedlightcurve}
\end{figure*}

We construct a systematic model ($S_{\mathrm{sys},i}(t)$) for each light curve by combining the principal components with the best-fitting weights. For the planet aperture, we calculate the corrected light curve as
\begin{equation}
    F_\mathrm{corrected, planet} = \left(F_i(t) / S_{\mathrm{sys},i}(t) - (1-c_i)F_\mathrm{star}(t)\right) / c_i. \label{eq:correction}
\end{equation}
We note that the random error term $\epsilon(t)$ is absorbed in $S_{\mathrm{sys},i}$ and sets a 0.1 to 0.3\% noise floor of the photometric time series. 
For comparison apertures, $c_i$ is close to zero. We calculate
\begin{equation}
    F_\mathrm{corrected, comparison} = F_i(t) / S_{\mathrm{sys},i}(t) / F_\mathrm{star}(t).
\end{equation}
In Figure~\ref{fig:correction_demo}, the bottom panel illustrates the F410M band planet aperture correction result as an example.
Figure~\ref{fig:correctedlightcurve} shows the corrected light curves in all apertures.
An accurate systematic model produces $F_\mathrm{corrected, planet} = F_\mathrm{planet}(t)$ plus random error. For comparison apertures, it produces a flat line with random fluctuation.

We compute the reduced $\chi^2$ values against a flat line to evaluate whether the light curves deviate significantly from a non-variable model. We assume each photometric measurement has identical uncertainty following a normal distribution. We estimate the photometric noise empirically from the root mean square of a high-pass-filtered light curve. The high-pass filter subtracts a boxcar-smoothed version (window size = 50 time points) from the original light curve. All light curves extracted from comparison aperture have reduced $\chi^2$ close to 1.0, and  none of these comparison light curves show $>3\sigma$ evidence for deviation from a flat line. In contrast, both $\beta$ Pic b light curves show strong evidence ($\gg 5\sigma$) of variability. The F210M light curve has reduced $\chi^2 = 1.26$ ($p=4.0 \times 10^{-14}$). The F410M light curve has reduced $\chi^2 = 1.51$ ($p=1.4 \times 10^{-43}$).

\begin{figure*}
    \centering
    \includegraphics[width=1\linewidth]{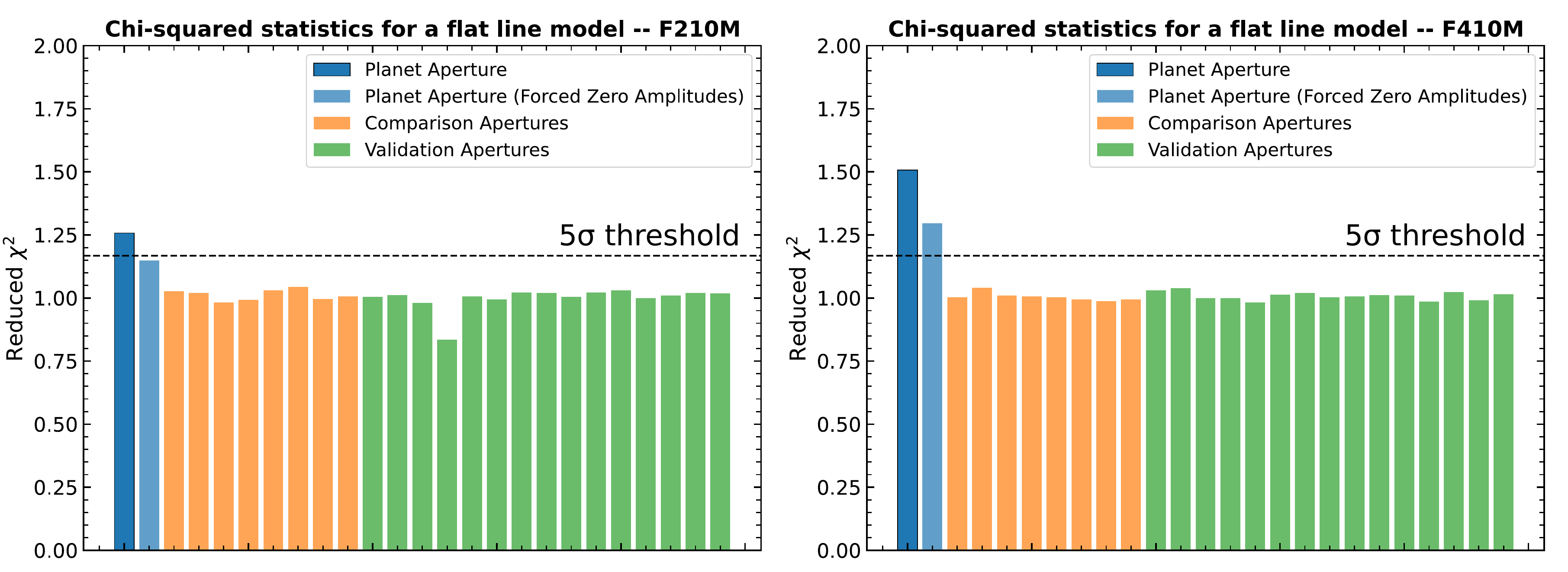}
    \caption{Significance of the variability detection demonstrated by $\chi^2$ statistics. We compare the corrected light curves extracted from the planet, comparison, and validation apertures with a flat line and compute the $\chi^2$ deviation. The planet's light curves show deviations far exceed a $5\sigma$ threshold. All  validation light curves are consistent with the flat line model.}
    \label{fig:significance}
\end{figure*}

\subsection{Validating the PCA method}

We examine whether the PCA correction generalizes beyond the apertures used to derive it. The systematic correction is constructed from the comparison apertures, so applying it back to light curves extracted from those same apertures is not an independent test. Flattened comparison light curves, shown in Figure~\ref{fig:correctedlightcurve}, are expected. The top three principal components explain over 95\% of the variance in the comparison aperture light curves. The principal component subspace and the comparison aperture light curve space therefore largely overlap.
We define fifteen validation apertures (marked by white dashed lines in Figure~\ref{fig:apertures}) whose light curves play no role in building the PCA systematic model. If the correction flattens these light curves, it demonstrates that the systematic model captures trends common across the field. 

The validation apertures at diverse positions sample a broad range of systematic environments. Six apertures lie at slightly smaller angular separations than the planet aperture (400 mas), three at slightly larger separations (650 mas), and six in distant background regions (1300 mas). Together they sample bright speckles, significant brightness gradients, and low-flux background. We extract light curves from these apertures and fit the same model (Equation~\ref{eq:S1}), holding the principal component curves fixed from the planet aperture analysis. Although the validation apertures contain negligible planet flux, we treat $c_i$ as a free parameter rather than fixing it to zero. This makes the test more conservative: the model is permitted to fit spurious variability as a planetary signal, allowing false positives to occur if the systematic correction is inadequate. We verify that fixing $c_i = 0$ yields consistent results, confirming that the conclusions do not depend on this choice.

We compare the corrected validation light curves to a flat line using a $\chi^2$ test. Deviation from a flat line indicates the PCA method is incomplete. The observed variability could then be residual systematics. If all validation light curves are consistent with flat lines, the PCA model successfully captures the systematic noise across diverse detector areas and conditions. The observed planetary variability cannot then be residual systematics.

Figure~\ref{fig:significance} shows the reduced $\chi^2$ values between validation light curves and flat lines as green bars. All validation light curves are consistent with a flat line. The planetary light curves continue to show significant deviation from a flat line. This shows the PCA model successfully removes systematic trends. The observed planetary variability cannot be residual systematics. This result validates the PCA method in effectively calibrating the high-contrast light curves.

\subsection{Identifying the periodic signals}

\begin{figure*}[th]
    \centering
    \includegraphics[width=1\linewidth]{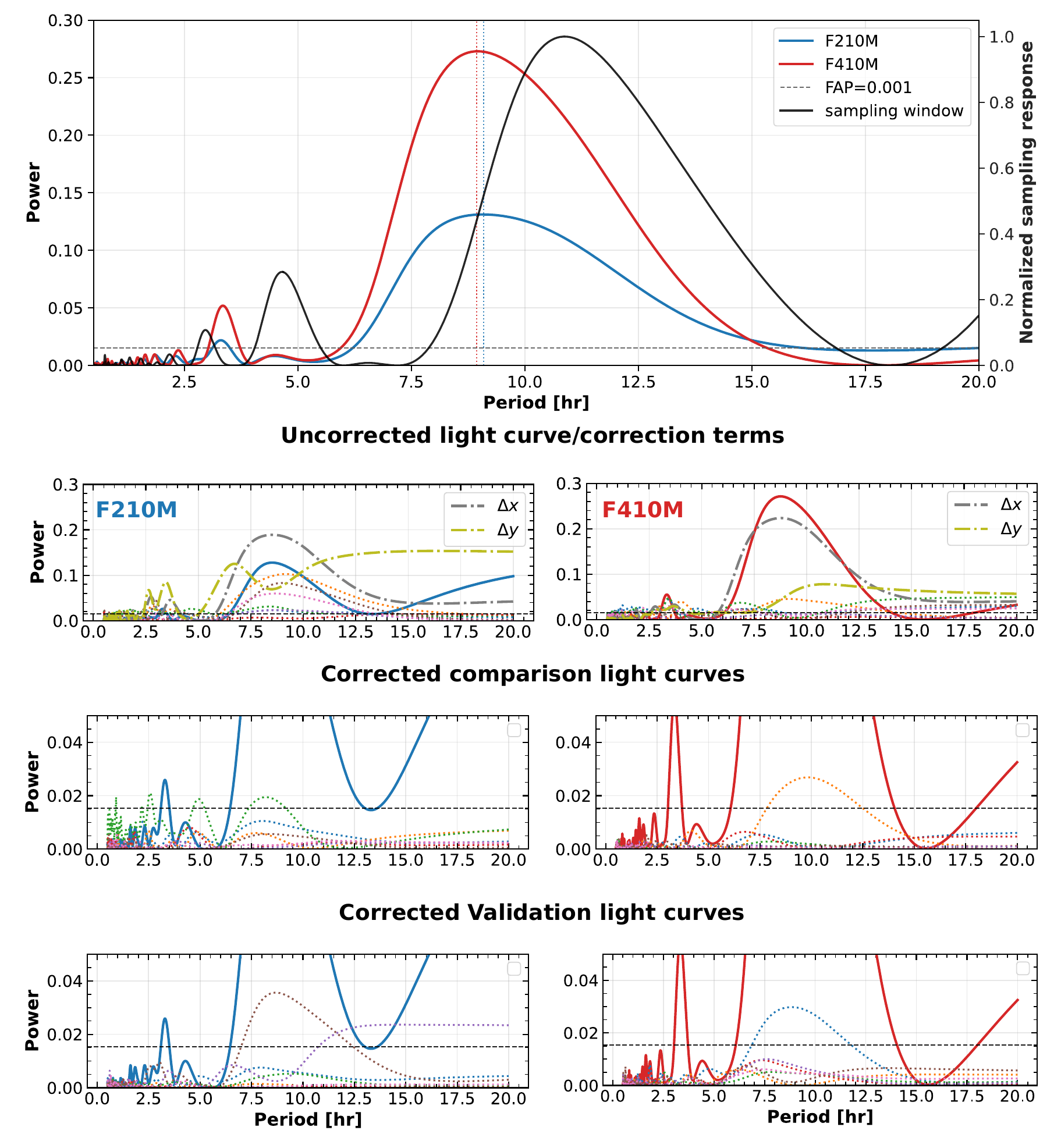}
\caption{Lomb-Scargle periodograms of $\beta$~Pic b in F210M (blue) and F410M (red).
\textit{Top:} Periodograms of the corrected target light curves. Both filters show a dominant peak near $\sim$8--9~hr above the 0.1\% false-alarm probability level (dashed line). Vertical dotted lines mark the peak period in each filter. The black curve shows the normalized sampling window, computed from the observation times and plotted against the right axis. The F410M light curve shows the stronger detection.
\textit{Second row:} Periodograms of the uncorrected target light curves (dotted lines) and centroid offset correction terms $\Delta x$ (gray dash-dot) and $\Delta y$ (olive dash-dot). These panels show periodic signatures associated with instrumental systematics and correction terms.
\textit{Bottom two rows:} Periodograms of corrected comparison light curves (third row) and corrected validation light curves (fourth row) from the same datasets.}
    \label{fig:lombscargle}
\end{figure*}

\begin{figure*}
    \centering
    \includegraphics[width=1\linewidth]{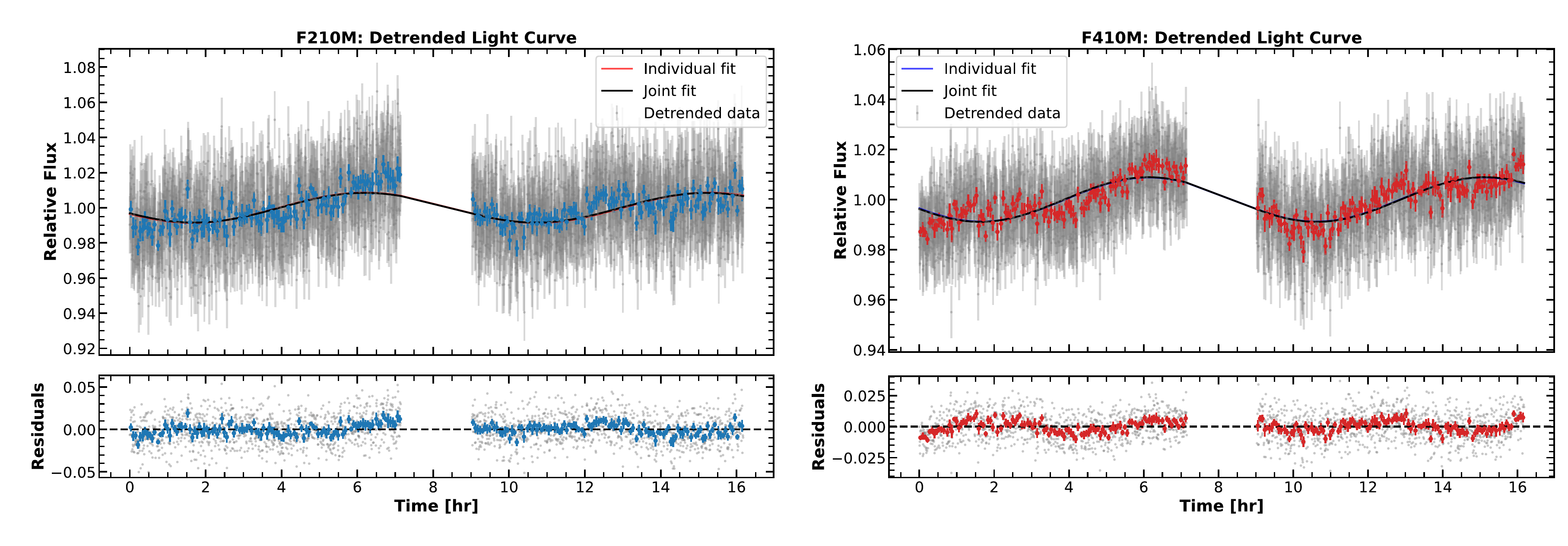}
    \caption{Detrended light curves of \bpicb{} in F210M (blue) and F410M (red) and the best-fitting sinusoids. The colored solid lines show the best-fit sinusoid from the individual-filter fit, and the black solid line shows the best-fit sinusoid from the joint fit across both filters. These two fits are almost identical and result in nearly indistinguishable solution. Residuals after subtracting the joint fit are shown in the bottom sub-panels.} \label{fig:sinusoid}
\end{figure*}

We searched for periodic signals in the detrended F210M and F410M light curves using the Lomb-Scargle periodogram \citep{Lomb1976, Scargle1982, VanderPlas2018}. The periodogram was computed over a period range of 0.5--20\,hr on a frequency grid uniform in frequency space with an oversampling factor of 100. We incorporated photometric uncertainties as weights and used the standard normalization \citep{Scargle1982}. We also computed the sampling window following \citet{VanderPlas2018}, by calculating the Lomb-Scargle periodogram of a constant light curve sampled at the observation times. The window depends only on the time sampling and shows where the observing cadence can imprint structure on the periodogram. False alarm probabilities (FAP) were estimated using the analytical approximation of \citet{Baluev2008} as implemented in \texttt{astropy} \citep{astropy2022}. The results are verified by a bootstrapping resampling approach implemented in \texttt{astropy} with $n=1000$ sampling points \citep{VanderPlas2018,astropy2022}. We applied the same periodogram analysis to the uncorrected target light curves and the centroid offset correction terms $\Delta x$ and $\Delta y$. We also analyzed the corrected comparison and validation aperture light curves.

Figure~\ref{fig:lombscargle} shows the results of our periodogram analysis. Both filters show a dominant peak well above the FAP\,=\,0.1\% threshold, at periods of 9.09\,hr (F210M) and 8.94\,hr (F410M). Light curves folded at these periods reveal coherent sinusoidal variability. The sampling window has a broad power spectrum, as expected for the two-block observing cadence, but its strongest peak is offset from the dominant $\beta$~Pic~b periodogram peak. Both periodograms additionally show marginally significant secondary peaks at 3.36\,hr exceeding the FAP\,=\,0.1\% threshold. Corrected light curves from the comparison and validation apertures show no significant periodic signals (Figure~\ref{fig:lombscargle}, bottom two rows). One comparison and one validation aperture in each band show a marginally significant peak, but with Lomb-Scargle powers below 30\% of the peak power observed in $\beta$~Pic~b (below 10\% in F410M). All remaining comparison and validation periodograms show no significant peaks. These results strongly support an astrophysical origin for the periodic signal in $\beta$~Pic~b, and disfavor instrumental systematics and post-processing artifacts.

A potential concern is that the dominant $\sim$8--9\,hr signal coincides in period with power seen in the $\Delta x$ centroid offset correction term (Figure~\ref{fig:lombscargle}, second row). The $\Delta x$ periodogram shows substantial power at periods comparable to the $\beta$ Pic b signals. This raises the question of whether the detected variability reflects a pointing-induced systematic rather than astrophysical flux modulation. We note, however, that the pointing drifts were included as part of the correction terms already applied to the detrended light curves (Equation~\ref{eq:systematics}). The persistence of the signal after this correction suggests it is not introduced by centroid drift.  Nonetheless, we perform multiple tests to address this concern further in Section~\ref{sec:validation}.

To verify the periodogram detection, we fit a sinusoidal model $F(t) = A\sin(2\pi t/P + \phi) + C$ directly to the detrended light curves using MCMC implemented in \texttt{emcee}. The quoted uncertainties are the formal posterior uncertainties from this sinusoidal model, derived from the 16th and 84th percentiles of the posterior distributions. They quantify the period precision under the assumed sinusoidal light-curve model and measured photometric uncertainties. Fitting each filter independently yields $P = 9.08 \pm 0.24$\,hr and $A = 0.87 \pm 0.04$\% for F210M, and $P = 8.96 \pm 0.10$\,hr and $A = 0.88 \pm 0.03$\% for F410M; the two periods agree within their combined $1\sigma$ uncertainties. Figure~\ref{fig:sinusoid} shows the best-fitting model. In a joint fit where the period is shared between filters but the amplitude and phase vary independently, we obtain $P = 9.00 \pm 0.13$\,hr, with amplitudes of $0.85 \pm 0.07$\% (F210M) and $0.89 \pm 0.04$\% (F410M). These uncertainties do not include uncertainty from possible mismatch between a sinusoidal model and the true rotational waveform, which may be non-sinusoidal or evolve over the observing sequence. We discuss additional uncertainty associated with possible astrophysical light-curve shape mismatch in Section~\ref{sec:discussion_period}. The two band light curves do not show phase or shape differences, unlike those observed in later type variable brown dwarfs \citep[e.g.,][]{Biller2024, McCarthy2025, Chen2025, Oliveros-Gomez2026}.

\section{Verifying the variability detection} \label{sec:validation}

Three systematic uncertainties could contaminate the variability detection. First, assuming a sinusoid model for the planet may bias the detection. Second, the PCA correction may leave residual systematic noise in the planet aperture or introduce spurious variability, either of which could produce a false periodic detection. Third, the detected signal may depend on the specific aperture placement and the choice of PCA basis used to construct the systematic correction.  We conduct tests to validate the variability signal against these sources of contamination.

\subsection{Test 1: Examining the impact of light curve model assumptions} \label{sec:model_impact}

We examine possible bias from the assumed planet light curve model by repeating the analysis with a flat planet light curve. This forces the model to explain the observed variability using only the systematic principal components. We derive corrected planet light curves and calculate their $\chi^2$ values against a flat line using the procedures from the previous section.

The corrected light curves deviate from a flat line with $>5\sigma$ significance. The F210M light curve has reduced $\chi^2 = 1.15$ ($p = 4.7 \times 10^{-6}$). The F410M light curve has reduced $\chi^2 = 1.30$ ($p = 7.7 \times 10^{-18}$). Figure~\ref{fig:significance} compares these values (marked as ``Forced Zero Amplitudes") with those derived when allowing the planet light curve amplitude to vary freely. 

Forcing a flat planet model results in weaker variability signals, demonstrating that the systematic model can partially absorb the variability signal. The reduced $\chi^2$ values decrease toward unity under this constraint (Figure~\ref{fig:significance}), indicating the corrected light curves are more consistent with a flat line. This reflects a degeneracy between the two components. However, the systematic model cannot fully account for the observed trends. The corrected light curves remain inconsistent with a flat model at $>3\sigma$ significance for both F210M and F410M observations. For F410M, the residual signal exceeds $5\sigma$ even under the forced-flat constraint. The pointing shifts, despite demonstrating potentially confounding frequency-domain structure (Figure~\ref{fig:lombscargle}), do not eliminate the variability. This is consistent with an astrophysical origin of the signal.

\subsection{Test 2: Injecting-and-recovering of variability signals}

\begin{figure*}[!t]
    \centering
    \includegraphics[width=1\linewidth]{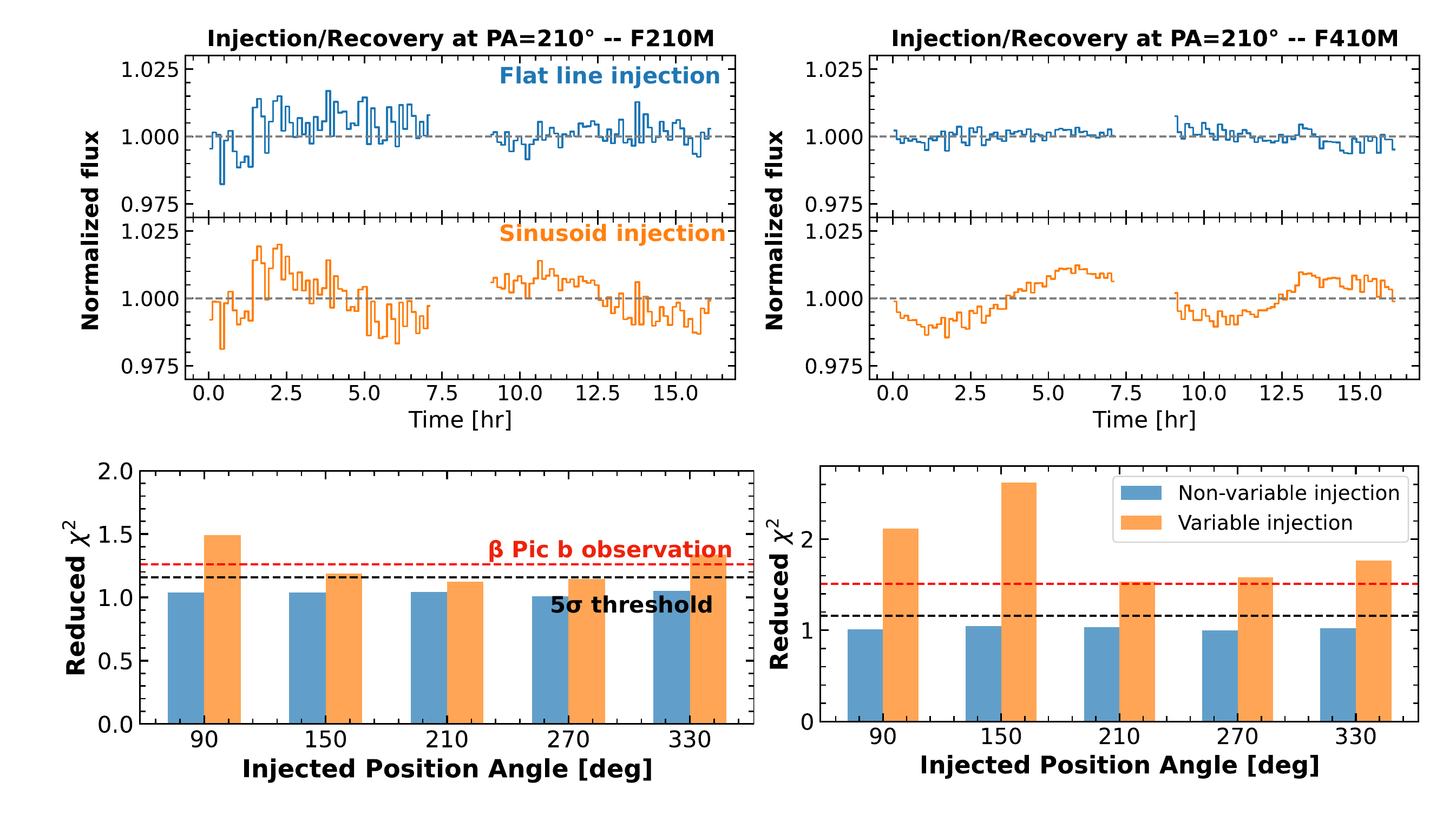}
    \caption{Summary of the injection and recovery test results for the F210M (left) and F410M (right). The upper panels show recovered light curves when injecting a flat line (blue) and a sinusoid (orange). The most noisy recovered light curves are presented to show the pessimistic scenario. The lower panels show the reduced $\chi^2$ between the recovered light curves and a flat line. The $5\sigma$ threshold and the observed values for $\beta$ Pic b are shown as black and red horizontal lines for comparison.}
    \label{fig:InjectionRecovery}
\end{figure*}

We perform injection-and-recovery tests to verify that our procedures recover a known light curve signal. We create simulated data by adding \texttt{stpsf} models to the centroid-aligned \texttt{calints.fits} data cube. The \texttt{stpsf} models are identical to the best-fit forward modeling PSFs. PSF evolution over the observation is not detectable, so this introduces negligible bias. We further assess robustness by injecting with random positional offsets of  $3\times$, and $10\times$ the measured centroid uncertainty (presented in Figure~\ref{fig:alignment}), each performed as an independent test, with results discussed below.

The test includes five injection positions. Each one constitutes an independent case. The injection positions share the same angular separation from the stellar centroid as $\beta$ Pic b (540~mas) but have different position angles: 90, 150, 210, 270, and 330 degrees. We consider both a flat and a variable injection model for each position angle. The variable model is a sinusoid with 1\% amplitude, 8.5~hr period, and a random phase  (uniformly distributed between $-\pi$ and $\pi$). This signal is similar to the observed variability in $\beta$ Pic b. Across both bands, we generate twenty simulated datasets.

We extract and correct the light curves using the same procedures as for the planetary light curves. The comparison aperture positions remain unchanged. If a comparison aperture overlaps with the injection aperture, we remove it from the comparison set. We compare the extracted light curves with the injection models using a $\chi^2$ test to determine whether the recovery deviates from the injection.

Figure~\ref{fig:InjectionRecovery} summarizes the injection-and-recovery test results. All flat line injections in both bands are recovered as flat light curves. This confirms that the three principal components sufficiently explain systematic trends. The systematic correction step does not generate false positive variability signals.

All sinusoidal injection recoveries favor a sinusoid over a flat line, with $\Delta$-BIC values ranging from a few hundred (F210M) to over one thousand (F410M). The recovered variability significance varies based on $\chi^2$ statistics. For F210M, three of five sinusoid recoveries show $>5\sigma$ deviation from a flat line. The other two cases have deviations with marginal significance. Example light curves in Figure~\ref{fig:InjectionRecovery} show these pessimistic scenarios. For F410M, all sinusoid recoveries deviate from a flat line exceeding the $5\sigma$ threshold. The variation in recovered significance arises from differences in systematic trends across detector locations and varying degeneracy between systematic trends and injected signals. Despite this degeneracy, all injected signals are accurately recovered. The reduced $\chi^2$ values from the injection-and-recovery tests bracket those observed in $\beta$ Pic b. The consistent $\chi^2$ values confirm the detection and statistical significance of the $\beta$ Pic b variability signal. The injection-and-recovery tests with added positional noise yield nearly identical results to those presented in Figure~\ref{fig:InjectionRecovery}. This confirms that at JWST's pointing stability level, positional uncertainty does not introduce detectable systematics into the light curves.

\subsection{Test 3: Pixel-level PCA}

\begin{figure}
    \centering
    \includegraphics[width=1\linewidth]{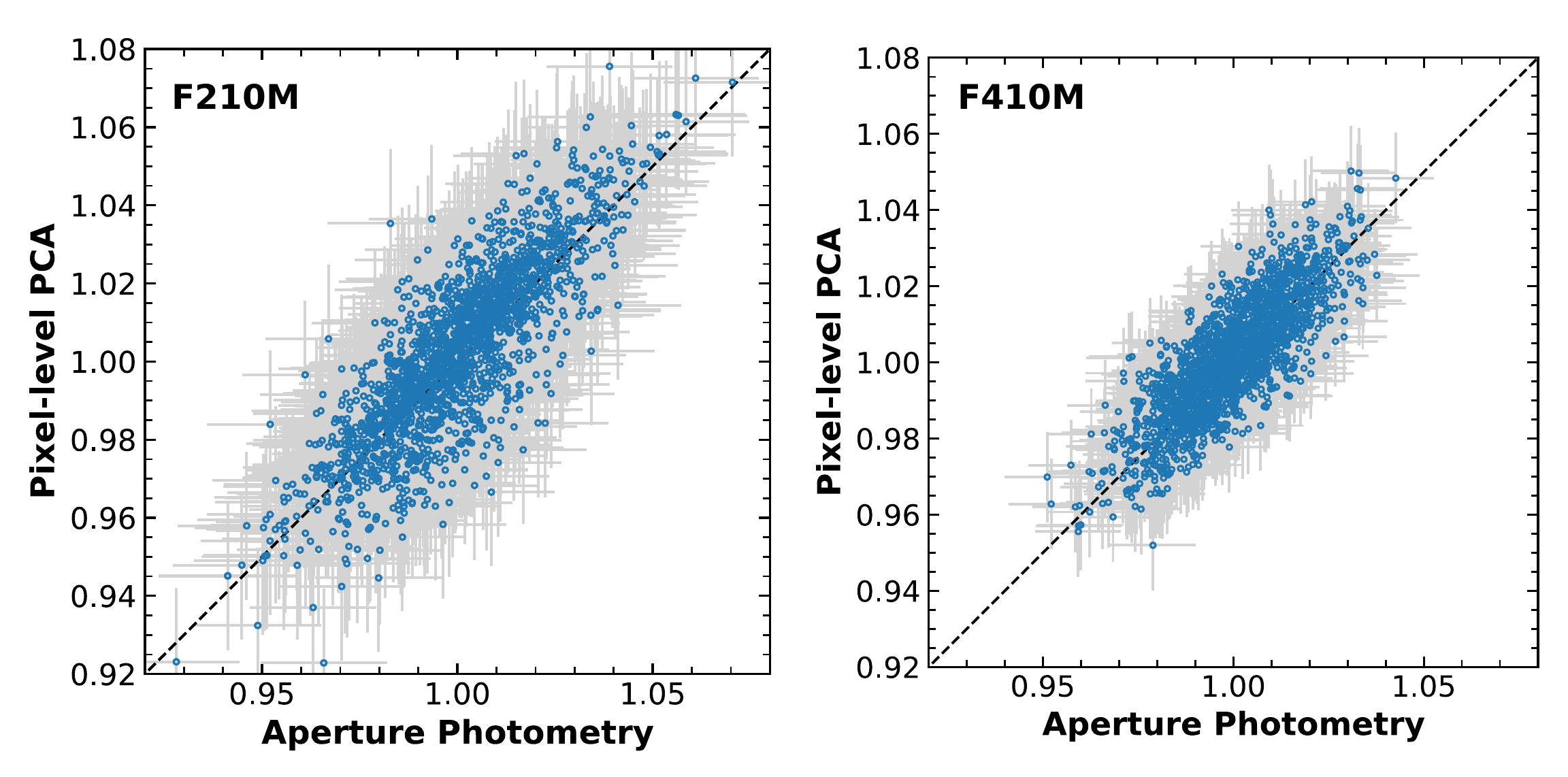}
    \caption{Comparisons between the nominal method and the pixel-level PCA method. The two measurements are shown in the $x$ and $y$ dimensions. The black dashed lines show the one-to-one correlation. Both methods produce statistically consistent results for F210M and F410M data.}
    \label{fig:pixelevelPCA}
\end{figure}

We perform PCA systematic correction on pixel-level light curves. The pixel-level PCA procedure is nearly identical to the nominal method described previously. The difference is that we apply the model in Equation~\ref{eq:S1} directly to pixel-level data before performing aperture integration. This removes systematics at the pixel level and avoids potential biases in aperture selection and placement. We use this method to verify the accuracy of the light curves.

We limit the calculation to a $3.6''\times3.6''$ box centered on $\beta$ Pic ($120\times120$ pixels in F210M and $60\times60$ pixels in F410M). For each filter and roll, we perform a PCA and select the first three principal components for systematic trend modeling. Fitting Equation~\ref{eq:S1} to each pixel-level light curve finds the best systematic model pixel by pixel. This systematic model is then inserted into Equation~\ref{eq:correction} to derive the corrected planetary light curve pixel by pixel. Only pixels near the planet PSF centroid contain meaningful signals. As the final step, we perform aperture integration ($r=3.5$ pixels for F210M and $r=2.0$ pixels for F410M, centered on the planet) to derive the planetary light curves.

Figure~\ref{fig:pixelevelPCA} compares light curves measured by the pixel-level PCA and nominal methods. Both methods agree fully for the F210M and F410M data, with $\chi^2_\mathrm{red} = \sum_i (F_{\mathrm{corrected,\,planet}, i} - F_{\mathrm{PCA,\,planet}, i})^2/\sigma_{\mathrm{planet},i}^2$ = 0.83 and 0.92 for the F210M and F410M results, respectively. This consistency confirms that the detected variability in $\beta$ Pic b is robust against aperture selection, the PCA basis, and the order of PCA and aperture integration.

\subsection{Summary of the test results}

These three tests confirm the statistical significance and astrophysical origin of the variability signals observed in $\beta$ Pic b. The light curves measured at the $\beta$ Pic b position show distinctively different behavior from light curves measured at other locations (Tests 1 and 3). Systematic trends are moderate and can be completely corrected using PCA (Tests 2 and 3). The PCA correction does not introduce false positive signals (Test 2). Considering all three tests, we assess the F210M variability detection as moderately strong ($\sim5\sigma$) and the F410M detection as decisive ($\gg5\sigma$). Because the variability patterns in both bands are nearly identical, the observed variability most likely originates from the planet itself rather than instrument- or data-reduction related systematics.

\section{Discussion}

\subsection{Interpreting the variability signals}\label{sec:discussion_period}

The tests presented in Section~4 confirm the astrophysical origin of the observed variability in $\beta$~Pic~b. The nearly identical periodicity recovered independently in the F210M and F410M light curves provides further support for this interpretation (Figure~\ref{fig:lombscargle}). Heterogeneous or temporally evolving atmospheric structures are the most natural explanation for photometric modulations that are coherent across two widely separated wavelength bands.

Longitudinal variations in clouds, chemical composition, and thermal structure drive rotational modulations in the disk-integrated light curves of solar system giant planets \citep[e.g.,][]{Karalidi2015,Simon2016,Ge2019}. Heterogeneous cloud coverage is generally considered as the primary driver of photometric and spectroscopic variability in brown dwarfs 
and planetary-mass objects \citep[e.g.,][]{Apai2013,Biller2024,McCarthy2025}. Under this interpretation, the periodic signal directly measures the planetary rotation period. The joint sinusoidal fit to the F210M and F410M light curves yields $P_{\rm rot} = 9.00 \pm 0.13$~hr. The formal uncertainty of $\pm0.13$~hr reflects only the statistical error from the sinusoidal fit. It does not account for the effects of atmospheric evolution, and is therefore likely underestimated. Precise light curve monitoring of brown dwarfs frequently reveals rapid evolution on timescales comparable to a single rotation period \citep[e.g.,][]{Apai2017,Zhou2022,Biller2024,Chen2025,McCarthy2025}. General circulation models of young, hot giant planets predict that high internal heat flux reduces radiative timescales and drives rapid reorganization of cloud structures \citep[]{Tan2021a,Tan2021b,Tan2025}. Additionally, light curves of brown dwarfs and solar system gas giants often exhibit double-peaked or non-sinusoidal morphologies \citep{Simon2016,Apai2017,Ge2019,Vos2018MNRAS.474.1041V,Biller2024}. These features introduce systematic uncertainties in the extraction of a unique rotation period from a single-epoch light curve.

\begin{figure}[th]
    \centering
    \includegraphics[width=1\linewidth]{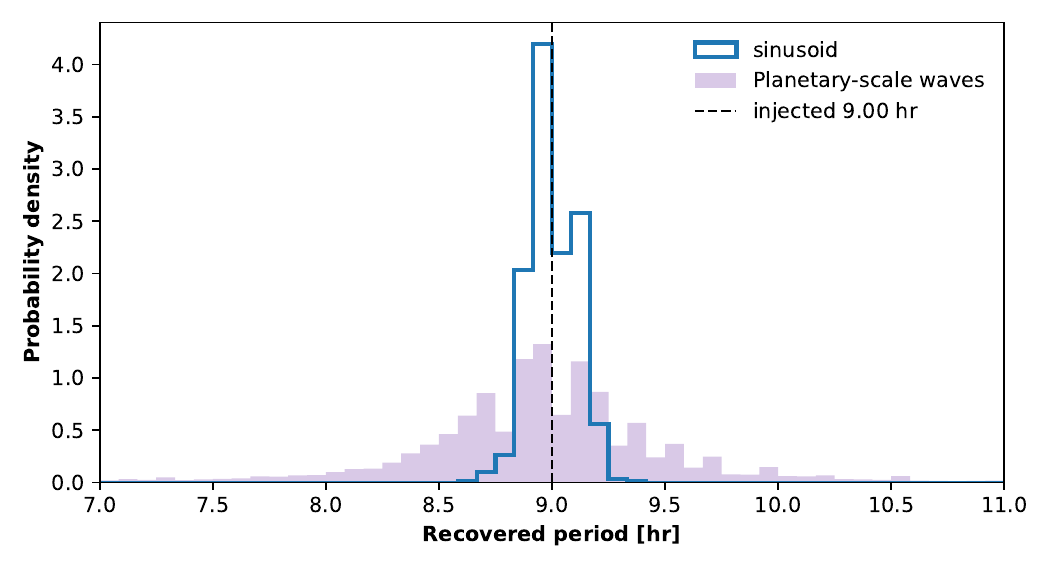}
    \caption{Recovery of the rotation period from synthetic F410M light curves sampled at the observed cadence and uncertainties. Blue shows period recovery of a sinusoid, and purple shows injections with $k=1$ and $k=2$ sinusoids motivated by the planetary scale wave models. The dashed line marks the injected period of 9.0\,hr. The sinusoidal cases recover a narrow period distribution, while the multi-sine waveforms remain centered near the injected period but broaden the 16th--84th percentile range to 8.5--9.4\,hr.}
    \label{fig:wave_recovery}
\end{figure}

We inject-and-recover plausible light curve models to estimate the effect of this model mismatch (Figure~\ref{fig:wave_recovery}). Synthetic light curves were sampled at the observed cadence and injected with a $\sim$9\,hr rotational signal and perturbed using the measured F410M photometric uncertainties. The non-sinusoidal injections followed the planetary scale wave model used by \citet{Fuda2024ApJ...965..182F} in fitting the TESS light curves of Luhman16AB. The model includes one or two components near the rotation period and one component near half the rotation period. The relative amplitude of the half-period component was drawn from 0.4--0.8 of the dominant rotational component, comparable to the fitted $k=2/k=1$ amplitudes in their Luhman~16B segments. We also allowed the amplitude to change linearly with time, motivated by the segment-to-segment amplitude evolution observed by \citet{Fuda2024ApJ...965..182F}.
The recovered periods of the planetary scale wave model injection remain centered near 9\,hr but with a broadened distribution that has a 16th--84th percentile range of 8.47--9.41\,hr. These tests support the $\sim$9\,hr rotation period, while showing that the formal statistical uncertainty can be underestimated due to inaccurate model assumptions.

We therefore adopt $P_{\rm rot} = 9.00 \pm 0.13$~hr as our best current measurement of the rotation period of $\beta$~Pic~b. Both the accuracy and precision of this value must be interpreted in the context of a dynamically active atmosphere. Additional monitoring epochs will be needed to disentangle atmospheric evolution from the underlying rotation signal and to improve the period determination.

\subsection{The planetary obliquity of $\beta$ Pic b}

\begin{figure*}[!th]
    \centering
    \includegraphics[width=1\linewidth]{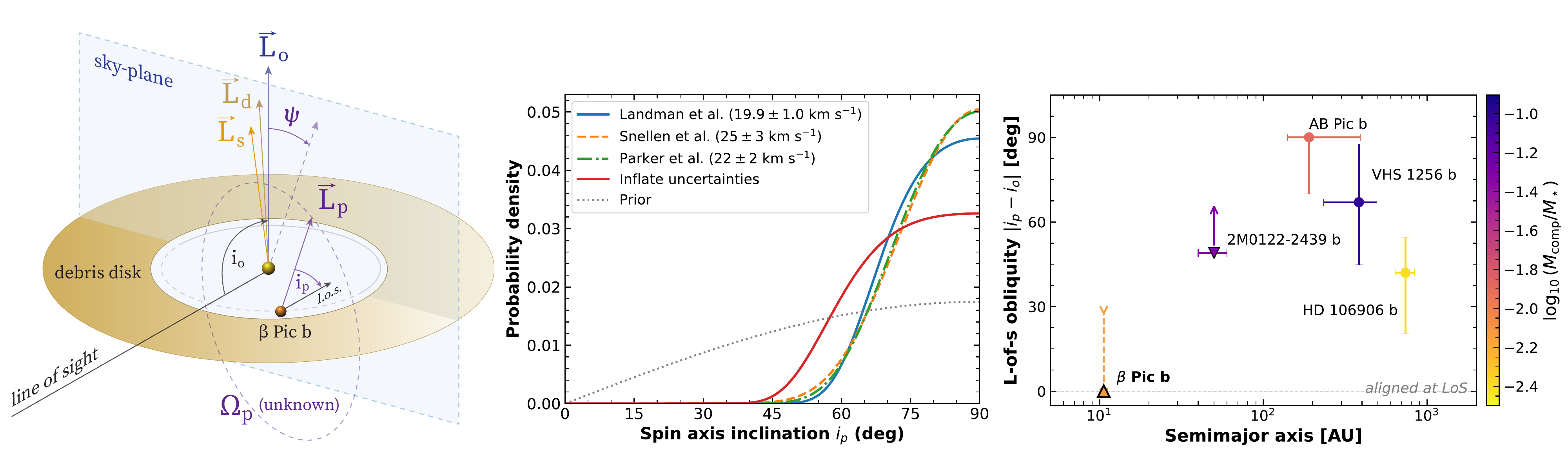}
    \caption{Angular momentum architecture and spin-axis inclination of $\beta$~Pic~b. \textit{Left:} The orbital ($\vec{L}_o$), disk ($\vec{L}_d$), and stellar spin ($\vec{L}_s$) angular momenta are mutually aligned to within a few degrees. This work constrains $i_p$ and $\vec{L}_p$ relative to the line of sight. \textit{Middle:} Posterior distributions of the spin-axis inclination $i_p$ of $\beta$~Pic~b derived from three independent $v \sin i$ measurements. One calculation assumed a $10\times$ inflated uncertainty in the period estimate. The dotted curve shows the random orientation prior (uniform in $\cos i_p$). All three posteriors peak at $i_p = 90\degr$, strongly favoring an equator-on viewing geometry. \textit{Right:} Line-of-sight obliquity $|i_p - i_o|$ versus semimajor axis for directly imaged companions with measured spin inclinations, color-coded by companion-to-host mass ratios.}
    \label{fig:obliquity}
\end{figure*}

We follow the method of \citet{Masuda2020AJ....159...81M} to combine the rotation period ($P_\mathrm{rot} = 9.00\pm0.13$~hr), the planetary radius, and the projected rotational velocity $v \sin i$ to constrain the line of sight spin-axis inclination of $\beta$~Pic~b. For the radius, we adopt a Gaussian prior $R\sim\mathcal{N}(1.4,\, 0.1)~R_\mathrm{Jup}$ from \citet{Landman2024A&A...682A..48L}, which encompasses recent spectrophotometric estimates \citep{Chilcote2017, GRAVITYCollaboration2020, Stolker2020}. Three $v \sin i$ measurements are available for $\beta$~Pic~b: $19.9 \pm 1.0$~km~s$^{-1}$ \citep{Landman2024A&A...682A..48L}, $22 \pm 2$~km~s$^{-1}$ \citep{Parker2024MNRAS.531.2356P}, and $25 \pm 3$~km~s$^{-1}$ \citep{Snellen2014Natur.509...63S}. We derive $i_p$ for each measurement independently and show the resulting posteriors in Figure~\ref{fig:obliquity}. The modes of all posteriors are $i_p = 90\degr$, strongly favoring an equator-on viewing geometry. Motivated by the waveform recovery tests in Section~\ref{sec:discussion_period}, we also test a calculation with an inflated period uncertainty by a factor of ten to account for uncertainty introduced by the unknown light-curve morphology.
Adopting the smallest available $v \sin i$ \citep{Landman2024A&A...682A..48L} and the 16th percentile of the posterior (as a proxy for $1\sigma$ bound) as a lower limit, we find $i_p > 62\degr$.

The spin-axis inclination $i_p \approx 90\degr$ is consistent with the  orbital inclination $i_\mathrm{orbit} = 89.04 \pm 0.03\degr$ \citep{GRAVITYCollaboration2020}. This shows no evidence for misalignment between the planetary spin axis and the orbital plane. We caution that the true three-dimensional obliquity $\psi$ cannot be constrained from the available data (Figure~\ref{fig:obliquity}, left panel), as it depends on the sky-plane position angle $\Omega_\mathrm{orbit} - \Omega_\mathrm{spin}$, where $\Omega_\mathrm{spin}$ for $\beta$ Pic b is not observable with current facilities, unlike the stellar spin axis \citep{Kraus2020}. Marginalizing over $\Omega_\mathrm{spin}$ uniformly leaves $\psi$ unconstrained. Nevertheless, the lower bound $|\,i_\mathrm{orbit} - i_p\,|$ is consistent with zero.

The obliquity constraint for $\beta$~Pic~b stands in stark contrast to all previously measured planetary obliquities: VHS~1256~b \citep{Poon2024AJ....168..270P}, 2M0122$-$2439~b \citep{Bryan2020AJ....159..181B}, HD~106906~b \citep{Bryan2021AJ....162..217B}, and AB~Pic~b \citep{Palma-Bifani2023, Gandhi2025} (Figure~\ref{fig:obliquity}, right panel). These wide-orbit companions collectively show a broad, roughly isotropic distribution consistent with top-down formation \citep{Poon2025}. In the $\beta$~Pic system, the stellar spin axis, the debris disk, the planetary orbital plane, and the planetary spin axis derived here are all mutually consistent with alignment \citep{Kraus2020}. Top-down formation via turbulent fragmentation imprints no preferred spin--orbit alignment, naturally producing large obliquities \citep{Jennings2021MNRAS.507.5187J, Offner2016}. Post-formation dynamical mechanisms can tilt an initially aligned planet to large obliquities, including planet--planet scattering \citep{li2021tilting}, Kozai--Lidov oscillations \citep{huang2023evolution}, secular spin--orbit resonances \citep[e.g.,][]{Ward2004AJ....128.2501W}, and giant impacts \citep[e.g.,][]{Slattery1992}. Neither class of mechanism is required by the observations of $\beta$~Pic~b. The absence of detectable line-of-sight obliquity thus provides independent dynamical support for a bottom-up formation pathway for $\beta$~Pic~b \citep{Johansen2010, batygin2018terminal}. 

$\beta$~Pic~b also resides in a system with a warped inner disk \citep{Mouillet1997,Heap2000} and a mildly eccentric orbit \citep{GRAVITYCollaboration2020, Lacour2021}, both of which remain difficult to attribute to a purely quiescent formation and evolution history. Whether the secular interaction between $\beta$~Pic~b and c \citep{Lagrange2019, Lagrange2020} can simultaneously account for the disk warp, the mild eccentricity of b's orbit, and the line-of-sight spin--orbit alignment reported here remains an open question.

\subsection{The origin of the variability signal}

\begin{figure*}[!th]
    \centering
    \includegraphics[width=1\linewidth]{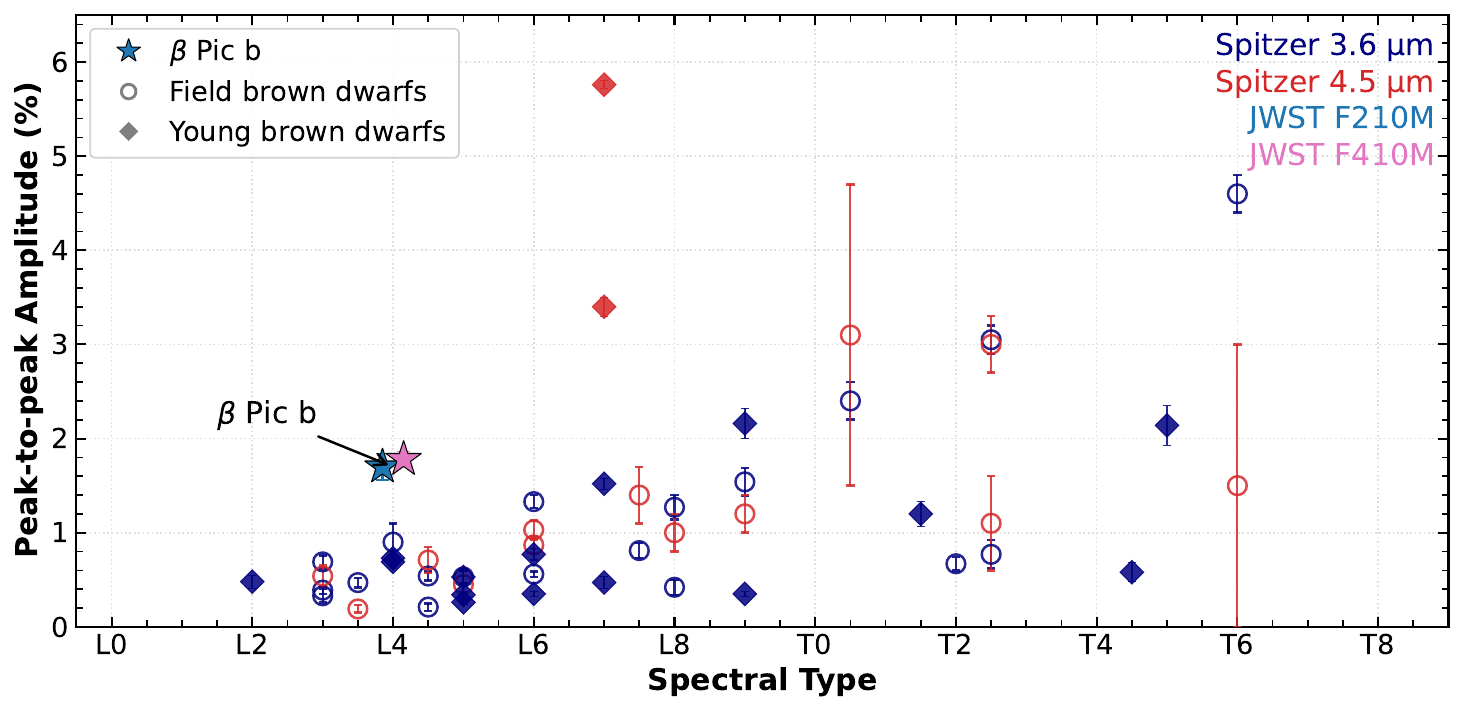}
    \caption{A comparison of the variability amplitude of \bpicb{} with those observed in brown dwarfs by Spitzer \citep{Metchev2015, Biller2018AJ....155...95B, Vos2022ApJ...924...68V, Zhou2020AJ....160...77Z}. Peak-to-peak amplitudes are adopted following literature convention. Color coding represents the observational bandpass: blue and pink for NIRCam F210M and F410M, respectively (this work); dark blue and red for Spitzer 3.6 \micron{} and 4.5 \micron{}, respectively. Open circles are field brown dwarfs; filled diamonds are young, low-gravity brown dwarfs.}    \label{fig:variability_comparison}
\end{figure*}

$\beta$~Pic~b has an early-L spectral type, placing it at the intersection of two distinct variability regimes in substellar objects. Late-M to early-L type brown dwarfs exhibit rotational modulations driven by magnetic starspots \citep[e.g.,][]{Scholz2015,Miles-Paez2019,Sutlieff2023MNRAS.520.4235S,OToole2026}. Simultaneously, this spectral type range marks the onset of silicate dust cloud formation, where heterogeneous cloud coverage becomes a dominant source of atmospheric variability \citep[]{Suarez2022,OToole2026}. Both mechanisms are viable drivers of the observed modulations in $\beta$~Pic~b.

Distinguishing between these mechanisms requires observations that probe different atmospheric pressure levels.  Figure~\ref{fig:observation} shows that the F210M and F410M bands probe nearly identical pressure levels in the atmosphere of $\beta$~Pic~b, which explains the near-identical variability amplitudes, phases, and light curve shapes recovered in the two bands ($0.85 \pm 0.07\%$ and $0.89 \pm 
0.04\%$, respectively). The diagnostic power of multi-band variability monitoring depends critically on whether the sampled pressure levels straddle the cloud base or regions of temperature profile variation \citep{Morley2014}, a behavior empirically confirmed by JWST spectroscopic light curves of brown dwarfs \citep{Biller2024,Chen2025, McCarthy2025,Oliveros-Gomez2026}. Because our two bands do not provide this leverage, the physical origin of the heterogeneity in $\beta$~Pic~b remains unconstrained by the present dataset.

Two future directions significantly advance the understanding of the driving mechanism. First, multi-epoch monitoring with additional photometric bandpasses that sample a wider range of pressure levels would break the degeneracy between magnetic and cloud-driven variability. Second, variable isolated brown dwarfs and planetary-mass objects with spectral types similar to $\beta$~Pic~b can serve as empirical references for interpreting the amplitude and wavelength dependence of the modulations.

To date, no early L-type brown dwarf has been monitored simultaneously at 2 and 4~$\mu$m with the photometric precision achievable by JWST. Spitzer photometric time series surveys of brown dwarfs offer the most informative comparison \citep{Metchev2015,Vos2017}. Figure~\ref{fig:variability_comparison} compares the variability amplitude of \bpicb{} with that of brown dwarfs observed by Spitzer at similar wavelengths \citep{Metchev2015, Biller2018AJ....155...95B, Vos2022ApJ...924...68V, Zhou2020AJ....160...77Z}. Among early-L objects (${\leq}$ L4), \bpicb{} exhibits the largest variability amplitude in this sample. This is consistent with its near-equator-on viewing geometry, as equator-on brown dwarfs and directly imaged planets preferentially show larger variability amplitudes \citep{Vos2017}. The equator-on geometry is also associated with enhanced silicate absorption at 8--11~\micron{}, where the column depth through the patchy cloud deck is maximized \citep{Suarez2022,Lam2026}; an SiO absorption feature has been tentatively identified in the $M$-band spectrum of \bpicb{} \citep{Parker2024MNRAS.531.2356P}. Follow-up campaigns combining ground-based spectrophotometric observations with the demonstrated capability of GRAVITY+ \citep{vonStauffenberg2026} and space-based photometric precision will help constrain the driving mechanism. Definitive characterization of the cloud structure would ideally require mid-infrared spectroscopic monitoring covering the silicate feature, but disk emission at these wavelengths renders such observations extremely challenging for \bpicb{} \citep{Worthen2024}. Spectroscopic variability monitoring of early-L brown dwarfs at 8--11~\micron{} offers a complementary and more accessible path to constraining the cloud properties responsible for the modulations we observe.

\subsection{Time-Series Coronagraphic Imaging with JWST}

This work demonstrates that JWST NIRCam coronagraphic imaging delivers high-precision time-series photometry of directly imaged planets. This was achieved through a novel PCA-based relative photometric method. The telescope's thermally stable environment, precise pointing, and high sensitivity beyond the coronagraph inner working angle together provide a photometric floor that ground-based adaptive optics systems cannot reach. Residual systematic noise is effectively removed by principal component analysis, and we have shown that the correction is robust against the choice of reference apertures and aperture size. In particular, these methods were effective in compensating for the discontinuity induced by a mirror tilt event partway through the time series. The achieved precision --- better than 0.5\% at 5-minute cadence --- exceeds the ground-based state of the art by more than a factor of 10 to 20 \citep{Apai2016ApJ...820...40A, Biller2021MNRAS.503..743B, Sutlieff2023MNRAS.520.4235S, Sutlieff2024MNRAS.531.2168S}. Future campaigns can further improve time-series coverage by adopting a single-roll design, eliminating the sampling gap introduced by telescope roll switching.

The demonstrated sensitivity is immediately applicable to a broad sample of directly imaged planets. NIRCam coronagraphic time series is well suited to measure rotation periods and map atmospheric heterogeneity in hot ($T_\mathrm{eff} \gtrsim 1000$~K), near-infrared bright companions. Applying this technique across the directly imaged planet population, in particular those with projected equatorial velocity readily available \citep{Hsu2026},  would build a statistical sample of planetary obliquities, establishing obliquity as a population-level diagnostic of formation and dynamical history \citep[e.g.,][]{Poon2026}. Beyond rotation and obliquity, the high-cadence light curves achieved here open additional science windows. The tentative spectroscopic variability reported by \citet{vonStauffenberg2026} with GRAVITY+ suggests that atmospheric evolution may already be detectable on short timescales and thus motivates follow-up observations. The photometric precision is sufficient to detect transits of Earth-sized moons around directly imaged giant planets \citep{Limbach2021,Wilson2025,Householder2025}. The temporal resolution also enables searches for high-frequency seismic oscillations that may trace past giant impacts. \citet{Zanazzi2025} showed that a Neptune-mass impact with \bpicb{} within the past $\sim$9–18 Myr would excite percent-level photometric oscillations with periods of $\sim$45 minutes; the photometric precision demonstrated here is sufficient to detect such a signal with a dedicated observing sequence. 

Extending this approach to longer wavelengths with MIRI would probe cooler atmospheric layers and enable variability studies of cold, wide-orbit companions inaccessible to NIRCam. More broadly, applying time-series coronagraphic imaging across a range of effective temperatures and masses strengthens the connection between atmospheric dynamics in brown dwarfs and giant planets, both within and beyond the solar system, by mapping heterogeneous atmospheres in objects with low internal heat flux. The precision and stability demonstrated here also set a strong foundation for time-resolved studies of progressively smaller companions, with rocky planets as a long-term goal.

\section{Conclusions}

We present the first detection of photometric variability in $\beta$~Pic~b. Our JWST NIRCam coronagraphic observations achieved photometric precision of 0.3\% over a continuous 16-hour baseline, enabling the detection of sub-percent variability amplitudes in a directly imaged planet for the first time. The key results of this work are as follows.

We developed and validated a framework for time-series photometry with JWST NIRCam coronagraphic imaging and demonstrated this framework on NIRCam observations of $\beta$ Pic b. The pipeline combines reference and angular differential imaging for PSF subtraction, principal component analysis for systematic noise removal, and injection-and-recovery tests for signal validation. Three independent tests confirm that the detected variability is astrophysical in origin and not a product of instrumental systematics or data reduction artifacts. This framework is directly transferable to future JWST time-series coronagraphic programs targeting other directly imaged planets and brown dwarf companions.

Both the F210M and F410M light curves of $\beta$~Pic~b show coherent, periodic variability at ${\sim}5\sigma$ for the F210M observations and $\gg5\sigma$ for the F410M observations significance. A joint sinusoidal fit yields a period of $P_{\rm rot} = 9.00 \pm 0.13$~hr and variability amplitudes of $0.85 \pm 0.07\%$ and $0.89 \pm 0.04\%$ in F210M and F410M, respectively.  The near-identical amplitudes and periods in the two bands confirm a common astrophysical origin. We interpret this period as the rotation period of $\beta$~Pic~b. The true uncertainty on this period measurement is likely larger than the formal fit error by a factor of a few (Figure~\ref{fig:wave_recovery}), given the potential for atmospheric evolution on timescales comparable to the rotation period.

Combining $P_{\rm rot}$ with the projected rotational velocity $v\sin i = 19.9 \pm 1.0$~km~s$^{-1}$ \citep[]{Landman2024A&A...682A..48L}, we derive the spin-axis inclination of $\beta$~Pic~b following \citet{Masuda2020AJ....159...81M}. The result strongly favors an equator-on viewing geometry. All major angular momentum vectors in the $\beta$~Pic system --- the planetary spin axis, the planetary orbit, the debris disk, and the stellar equator --- are consistent with alignment. This is in stark contrast to the large obliquities measured for wide-separation planetary-mass companions such as VHS~1256~b \citep{Poon2024AJ....168..270P}, which is consistent with forming top-down via gravitational fragmentation. The aligned line-of-sight-projected obliquity of $\beta$~Pic~b provides independent dynamical evidence for disk-mediated, bottom-up formation, and establishes obliquity as a discriminating observable between formation pathways.

This work demonstrates that time-series coronagraphic imaging with JWST is a powerful new observing mode for characterizing the rotation and atmospheric dynamics of directly imaged planets. The combination of space-based stability, infrared sensitivity, and coronagraphic contrast enables variability studies that were previously inaccessible. Applied to a broader sample of directly imaged planets and brown dwarf companions, this mode will constrain the spin distribution, obliquity architecture, and atmospheric structure of the directly imaged planet population.

\begin{acknowledgements}
We thank the referee for a constructive report. This work is based on observations made with the NASA/ESA/CSA James Webb Space Telescope. The data were obtained from the Mikulski Archive for Space Telescopes at the Space Telescope Science Institute, which is operated by the Association of Universities for Research in Astronomy, Inc., under NASA contract NAS 5-03127 for JWST. These observations are associated with program JWST-GO-4758. Support for this work was provided by NASA through a grant from the Space Telescope Science Institute under that program. JMV acknowledges support from a Royal Society - Research Ireland University Research Fellowship (URF/1/221932, RF/ERE/221108) and the European Union through the Exo-PEA ERC project (grant number 101164652). Views and opinions expressed are however those of the author(s) only and do not necessarily reflect those of the European Union or the European Research Council Executive Agency. Neither the European Union nor the granting authority can be held responsible for them. BJS acknowledges funding by the UK Science and Technology Facilities Council (STFC) grant nos. ST/V000594/1 and UKRI1196. S. Petrus's research was supported by an appointment to the NASA Postdoctoral Program at the NASA-Goddard Space Flight Center, administered by Oak Ridge Associated Universities under contract with NASA.
    All of the data presented in this article were obtained from the Mikulski Archive for Space Telescopes (MAST) at the Space Telescope Science Institute. The specific observations analyzed can be accessed via \dataset[doi:10.17909/9s0z-6s43]{https://doi.org/10.17909/9s0z-6s43}.

\end{acknowledgements}

\facilities{JWST (NIRCam)}

\software{spaceKLIP \citep{Carter2025},
          STPSF \citep{2012SPIE.8442E..3DP,2014SPIE.9143E..3XP},
          pyKLIP \citep{Wang2015pyklip},
          lmfit \citep{Newville2016},
          emcee \citep{ForemanMackey2013},
          george \citep{ForemanMackey2015},
          astropy \citep{astropy2022}
          }

\appendix
\section{Tilt Event and Micrometeor Flash Diagnostics}
\label{app:tilt_event}

As noted above, we found an abrupt discontinuity in the time series measurements partway through the first observation, affecting both the derived photometry of \bpicb{} and the inferred position of the host star relative to the coronagraph, which we infer was due to a mirror tilt event at that time. Occasional small abrupt changes in JWST's mirror alignment are a known property of the observatory, believed to arise due to the release of stored stresses from differential thermal contraction during JWST's post-launch cooldown to operational temperatures. Tilt events occurred relatively frequently early in the mission, and gradually less frequently over time as the stored stresses have dissipated. \citep{2024SPIE13092E..10T}

\begin{figure}
    \centering
    \includegraphics[width=1\linewidth]{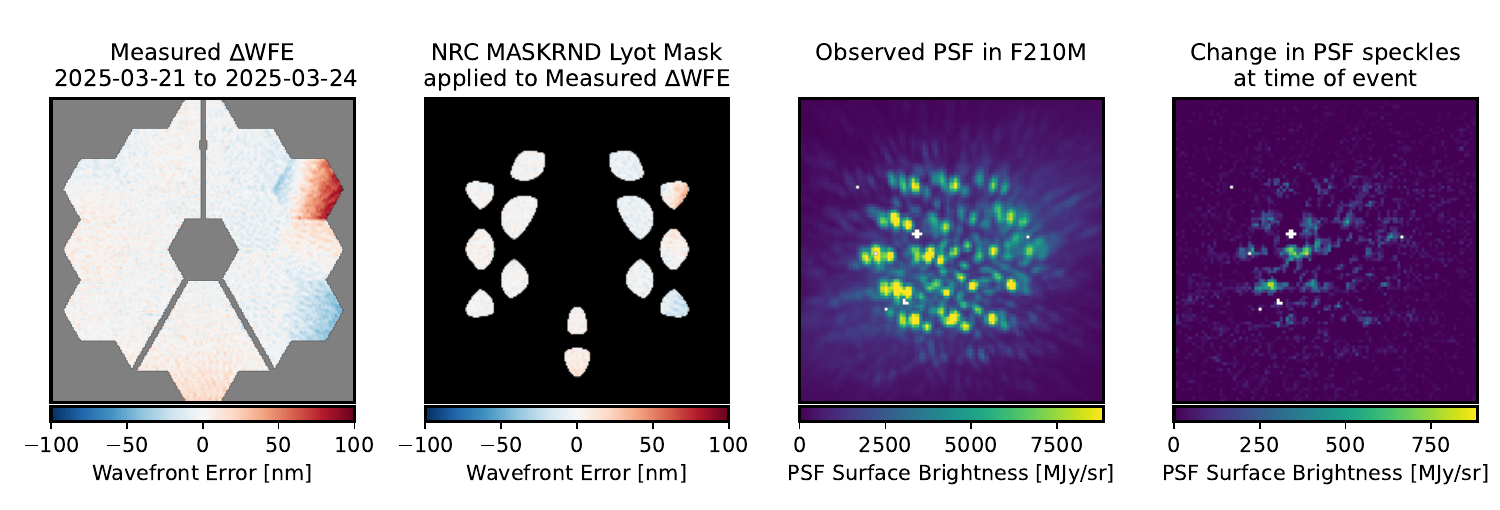}
    \caption{
    Observed mirror tilt and resulting coronagraphic PSF change. \emph{Left:} Measured $\Delta$WFE from WFS observations bracketing these data.Moderate tilts are observed affecting segments on the right (+V2) deployed wing of the primary. The observed pattern shows only piston-tip-tilt segment motions; no micrometeor impacts are detected. 
    \emph{Left center:} Same $\Delta$WFE multiplied by the NIRCam Lyot stop. The affected segments are mostly but not entirely blocked. Right center: Observed coronagraphic PSF of $\beta$ Pic in F210M; the debris disk is faintly visible as a diffuse diagonal without PSF subtraction. \emph{Right:} Change in PSF speckle pattern (after minus before) at the time of the discontinuity event. The display scale is $10\times$ smaller than the full PSF panel; the speckle change is subtle but sufficient to produce the discontinuities seen in the \bpicb\ time series.}
    \label{fig:mirrortilt}
\end{figure}

Wavefront sensing (WFS) observations directly measure JWST's mirror alignment periodically. Comparing the pair of WFS observations closest in time before and after this program's observations, specifically the WFS starting on 2025-03-21 03:29 and 2025-03-24 03:06 UTC, directly shows a change in alignment particularly affecting the segments on the right (+V2) deployable wing of the primary, and with a maximum change in wavefront of about 100 nm. As shown in Figure \ref{fig:mirrortilt}, the particular spatial pattern is recognizable as a wing hinge tilt event, a common type of tilt event affecting the hinge and latch mechanism supporting those three segments. Because WFS measurements are only taken every few days, typically the time of a tilt event during that period cannot be known with certainty, and the timing can only be inferred indirectly based on the circumstantial evidence of changes seen in the science data. 

The optical effect of a tilt event is, in general, a small change in PSF morphology, for instance a change in speckles in the PSF wings. For the specific case of NIRCam coronagraphy, the coronagraph's Lyot pupil mask blocks $\sim 85\%$ of the primary mirror. For this particular tilt event, it happens to be the case that the mirror regions most strongly affected by the tilt were blocked; as a result the change in the coronagraph PSF morphology is subtle, rather than dramatic. We verified that there is a detectable change in the coronagraphic PSF speckle pattern at the time of the event, specifically by comparing integrations 19 and 20 in file \texttt{jw04758001001\_03108\_00001\_nrca2\_calints.fits}. Furthermore that delta PSF is qualitatively broadly consistent with a simulated coronagraphic PSF difference made using STPSF with the before and after WFS measurements.  

Meanwhile, JWST's FGS would have seen the full area of the primary mirror, and the redistribution of light within the PSF due to the tilt event would slightly affect the measured PSF centroid. In other words, the FGS saw the full tilt $\Delta$WFE, while NIRCam mostly did not see it. Thus it acts like a step-function differential WFE between the two instruments, and as a result the FGS closed-loop tracking on the guide star led to an inadvertent tiny systematic shift of the science target star relative to the coronagraphic mask in NIRCam. This explains the discontinuity in measured star position of the host star. Furthermore, an inadvertent shift of this kind would tend to make the star less well aligned to the coronagraphic mask, hence increasing the amount of starlight leaking past the coronagraph. This explains the jump in the measured host star flux as shown in Figure 3. Together the change in PSF speckle pattern and host star alignment then naturally lead to the discontinuities in the extracted photometry light curves for \bpicb{} as shown in Figure\,\ref{fig:rawlightcurve}. Collectively these several lines of evidence support the inference that the mirror tilt sometime between 2025-03-21 03:29 and 2025-03-24 03:06 UTC must have happened at 2025-03-21 12:56 UTC, and thereby caused the observed discontinuities.

\begin{figure}
    \centering
    \includegraphics[width=1\linewidth]{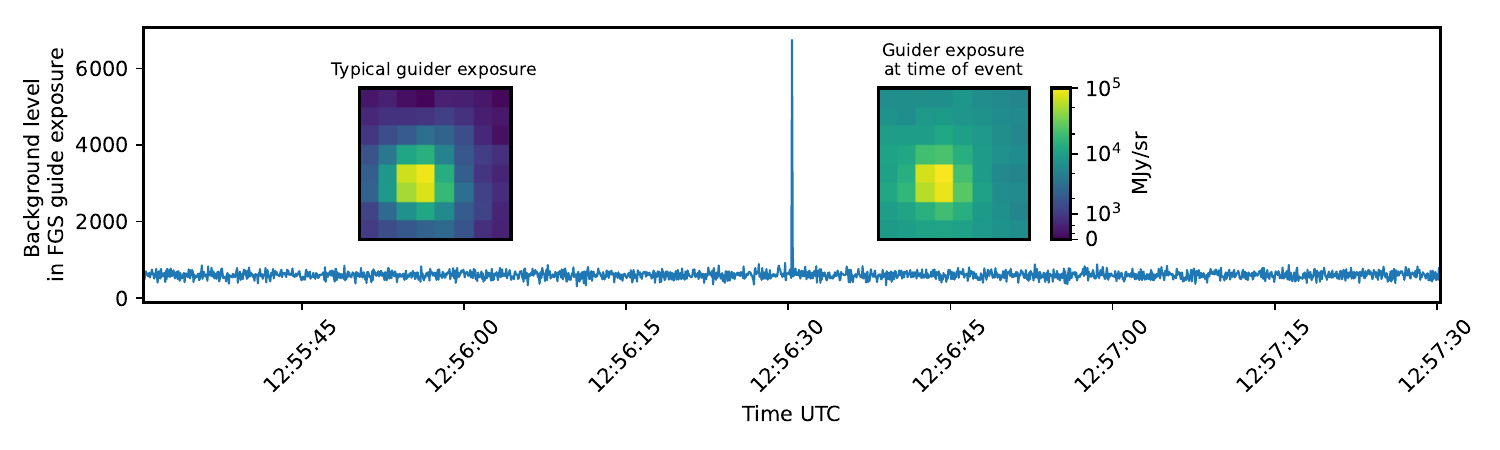}
    \caption{
        A flash observed by JWST's guider at the same time as the discontinuity in the time series data. The main panel shows the median background level in the $8\times8$ ``postage stamp'' guide exposures. The background spikes sharply at the moment of the PSF change. Insets compare a typical guide star image to the image at the time of the event. The flash morphology is consistent with a micrometeor strike. A comprehensive analysis of such events in FGS and science data will be presented in Telfer, Melendez, et al.\ (STScI technical report, submitted).}
    \label{fig:guider_flash}
\end{figure}

A second, independent line of evidence also directly indicates a perturbation to JWST's optics at that time: a brief flash of light seen in the guider, uniformly filling the subarray field of view for a single time sample. Analysis at STScI has detected such brief flashes seen in FGS data (and, rarely, in science data); their origin is now understood to be illumination by the brief pulse of thermal radiation when JWST is struck by a micrometeor.  These analyses and the properties of the flashes are described in detail in an upcoming paper (R. Telfer, M. Melendez, et al., 2026).  Such a flash was detected in the FGS guiding data at precisely 2025-03-21 12:56:30.349 UT $\pm 0.1$ seconds (the FGS reads out at 15.6 Hz during fine guide operations, and JWST's onboard absolute clock accuracy is 0.1 s or better) (Figure~\ref{fig:guider_flash}).  This precisely agrees with the time inferred from the change in the coronagraphic PSF. We note that the wavefront sensing data do \textit{not} show any indication of a micrometeor strike on any mirror during this time period; the impact appears to have been with some non-optical surface, plausibly the telescope structure or the stray-light baffle frill around the primary mirror. 

This combination of events provides, for the first time, an example of a bright flash occurring at the same time as a tilt event. The apparently random stochastic nature of tilt events has previously raised a natural question ``Why did a mirror tilt occur then, and not at some other time?'' For the first time this observation provides a ``smoking gun'' for at least some tilt events: a micrometeor impacted JWST's structure, probably somewhere near the wing hinge, and that brief impulse was sufficient to trigger a small stick-slip relaxation of the hinge by some handful of nanometers. The resulting small shifts in mirror positions led to distinct changes in the PSFs seen by NIRCam and the FGS, and the FGS simultaneously also directly saw the brief transient dissipation of thermal energy at the impact site.

\bibliography{BetaPicb_references}{}
\bibliographystyle{aasjournalv7}



\end{document}